\documentclass[conference]{sty/ndss}
\usepackage[square,comma,numbers,sort&compress]{natbib}

\usepackage{colortbl}

\usepackage{silence}
\usepackage{lipsum}
\usepackage[hyphens]{url}
\usepackage[breaklinks,colorlinks]{hyperref}
\usepackage[usenames,dvipsnames]{xcolor}
\hypersetup{citecolor=blue,linkcolor=blue}
\usepackage{amsmath,amsopn,amssymb}
\usepackage{endnotes,microtype,xspace,graphicx,fancyvrb,multirow}
\usepackage{booktabs}
\usepackage{array,underscore,relsize}
\usepackage[T1]{fontenc}
\usepackage{times}
\usepackage{fancyhdr,lastpage}
\let\labelindent\relax
\usepackage[linesnumbered,lined,ruled,noend]{algorithm2e}
\usepackage[labelfont=bf,font=small,skip=5pt]{caption}
\usepackage{fp}
\usepackage{siunitx}

\usepackage{balance}

\usepackage{subcaption} 
\usepackage{latexsym}
\usepackage[frozencache=true,cachedir=minted-cache]{minted}
\setminted{fontsize=\small, autogobble=true, breaklines}
\usepackage{tikz}
\newcommand*{\circled}[1]{\lower.7ex\hbox{\tikz\draw (0pt, 0pt)%
    circle (.5em) node {\makebox[1em][c]{\small #1}};}}

\usepackage{markdown}
\usepackage{bbding}

\markdownSetup{fencedCode = true}

\newcommand{\sys}{\textsc{BULKHEAD}\xspace}

\newcommand{\cc}[1]{\mbox{\smaller[0.5]\texttt{#1}}}

\fvset{fontsize=\scriptsize,xleftmargin=8pt,numbers=left,numbersep=5pt}

\input{code/fmt}

\def\Snospace~{\S{}}

\newcommand{\x}{$\times$\xspace}

\newif\ifdraft\drafttrue
\newif\ifnotes\notestrue
\ifdraft\else\notesfalse\fi

\input{glyphtounicode}
\newcolumntype{R}[1]{>{\raggedleft\let\newline\\\arraybackslash\hspace{0pt}}p{#1}}

\newcommand{\includepdf}[1]{
  \includegraphics[width=\columnwidth]{#1}
}

\newcommand{\squishlist}{
\begin{itemize}[noitemsep,nolistsep]
  \setlength{\itemsep}{-0pt}
}
\newcommand{\squishend}{
  \end{itemize}
}

\usepackage{tikz}

\usepackage{xstring}
\newcommand{\PP}[1]{
\vspace{2px}
{\it \IfEndWith{#1}{.}{#1}{#1.}}
}

\newcommand{\boxbeg}{
\vspace{2px}
\noindent\begin{tabular}{|l|}\hline
\begin{minipage}{3.2in}
\vspace{2px}
\noindent
}

\newcommand{\boxend}{
\vspace{2px}
\end{minipage}\\ \hline
\end{tabular}
\vspace{-10pt}
}

\usepackage{soul} 
\begin{document}

\title{BULKHEAD: Secure, Scalable, and Efficient Kernel Compartmentalization with PKS\vspace{-1em}}





\author{\IEEEauthorblockN{Yinggang Guo\IEEEauthorrefmark{1}\IEEEauthorrefmark{2},
Zicheng Wang\IEEEauthorrefmark{1},
Weiheng Bai\IEEEauthorrefmark{2}, 
Qingkai Zeng\IEEEauthorrefmark{1} and
Kangjie Lu\IEEEauthorrefmark{2}}
\IEEEauthorblockA{\mbox{\IEEEauthorrefmark{1}State Key Laboratory for Novel Software Technology, Nanjing University}, \IEEEauthorrefmark{2}University of Minnesota}
\{gyg, wzc\}@smail.nju.edu.cn, bai00093@umn.edu, zqk@nju.edu.cn, kjlu@umn.edu}


\IEEEoverridecommandlockouts
\makeatletter\def\@IEEEpubidpullup{6.5\baselineskip}\makeatother
\IEEEpubid{\parbox{\columnwidth}{
		Network and Distributed System Security (NDSS) Symposium 2025\\
		23 - 28 February 2025, San Diego, CA, USA\\
		ISBN 979-8-9894372-8-3\\
            https://dx.doi.org/10.14722/ndss.2025.23328\\
            www.ndss-symposium.org
}
\hspace{\columnsep}\makebox[\columnwidth]{}}

\maketitle


\begin{abstract}
The endless stream of vulnerabilities urgently calls for principled mitigation to confine the effect of exploitation. However, the monolithic architecture of commodity OS kernels, like the Linux kernel, allows an attacker to compromise the entire system by exploiting a vulnerability in any kernel component. Kernel compartmentalization is a promising approach that follows the least-privilege principle. However, existing mechanisms struggle with the trade-off on security, scalability, and performance, given the challenges stemming from mutual untrustworthiness among numerous and complex components.

In this paper, we present \sys, a secure, scalable, and efficient kernel compartmentalization technique that offers bi-directional isolation for unlimited compartments. It leverages Intel's new hardware feature PKS to isolate data and code into mutually untrusted compartments and benefits from its fast compartment switching. With untrust in mind, \sys introduces a lightweight in-kernel monitor that enforces multiple important security invariants, including data integrity, execute-only memory, and compartment interface integrity. In addition, it provides a locality-aware two-level scheme that scales to unlimited compartments. We implement a prototype system on Linux v6.1 to compartmentalize loadable kernel modules (LKMs). Extensive evaluation confirms the effectiveness of our approach. As the system-wide impacts, \sys incurs an average performance overhead of 2.44\% for real-world applications with 160 compartmentalized LKMs. While focusing on a specific compartment, ApacheBench tests on \cc{ipv6} show an overhead of less than 2\%. Moreover, the performance is almost unaffected by the number of compartments, which makes it highly scalable.

\end{abstract}

\section{Introduction}
\label{sec1}

The Operating System (OS) kernel serves as the cornerstone of system software, with expanding functionality for a diverse range of user programs and hardware devices. Despite its significance, the huge codebase which is written in unsafe languages faces a continual influx of vulnerabilities. Illustratively, the Linux kernel, a widely utilized OS kernel, has witnessed a stark increase in Common Vulnerabilities and Exposures (CVE) reporting. In 2023, the incidence of CVEs surged by over 179\% compared to 2013, cumulatively amounting to 2,814 CVEs over the decade~\cite{CVE23Linux}. 

Given that complete elimination of vulnerabilities remains an elusive goal, confining the impact of each vulnerability, including those yet undetected, is essential for system security. However, the upsetting fact is that most mainstream OS kernels, like the Linux kernel, are monolithic for performance and compatibility reasons, thus lack fault isolation. All kernel components share the supervisor privileges for data access and code execution, most of which are irrelevant to their intended duties. As a result, exploiting a single vulnerability in any component can potentially compromise the entire system. 

A viable solution to this pervasive issue is kernel compartmentalization~\cite{mckee2022preventing, roessler2021muscope}, which offers principled and systematic protection against vulnerabilities through the principle of least privilege~\cite{saltzer1975protection}. In this approach, kernel modules are isolated into separate compartments, with each compartment having access only to the data and code necessary for its functionality. Consequently, the impact of an exploited module is constrained, preventing systemic damage. 

In the face of various vulnerabilities and powerful attackers, the kernel compartmentalization mechanism must achieve the trifecta of security, scalability, and performance. Particularly, in order to strictly confine any exploitation within compartment boundaries, the mechanism should guarantee not only data access but also control flow transfer. Besides, compartment interfaces must be well-protected against confused deputy attacks~\cite{Lefeuvre2023}, which confuse a compartment to perform sensitive operations on behalf of the untrusted caller with malicious inputs. Scalability is another significant objective. Given the huge codebase of the OS kernel, fine-grained compartmentalization calls for a large number of isolated compartments. Finally, developers constantly trade off the benefits of protection against performance overhead. Only efficient mechanisms will bolster wide adoption in practice.

Unfortunately, existing works fail to fulfill these desirable objectives. Microkernels~\cite{microkernel95, klein2009sel4, gu2020harmonizing, redleaf20} minimize the attack surface by moving most kernel components into isolated user processes. Although providing strong protection, heavy inter-process communication (IPC) leads to low performance~\cite{gu2020harmonizing}. The complete redesign of the OS also hinders its general application. Software fault isolation (SFI)-based approaches~\cite{sfi93, xfi06, bgi09, mao2011software} establish an isolated domain by instrumenting each memory access instruction with security checks during compilation time, which results in significant overhead. Virtualization-based approaches~\cite{xiong2011huko, virtuos13, secpod15, lxd19, lvd20, huang2022ksplit} employ a hypervisor to protect execution domains and can form different memory views with the extended page tables (EPTs). However, the additional layer makes the Trusted Computing Base (TCB) more complex. Running systems in virtual machines (VMs) results in extra overhead and nested virtualization restrictions. There are also some efforts~\cite{dautenhahn2015nested, cho2017dynamic, MANES2018130DIKernel, iskios21} that utilize various hardware features to protect the kernel but cannot scale to multiple compartments because of the hardware limitations.

\begin{figure}
\begin{center}
\includegraphics[width=.82\linewidth]{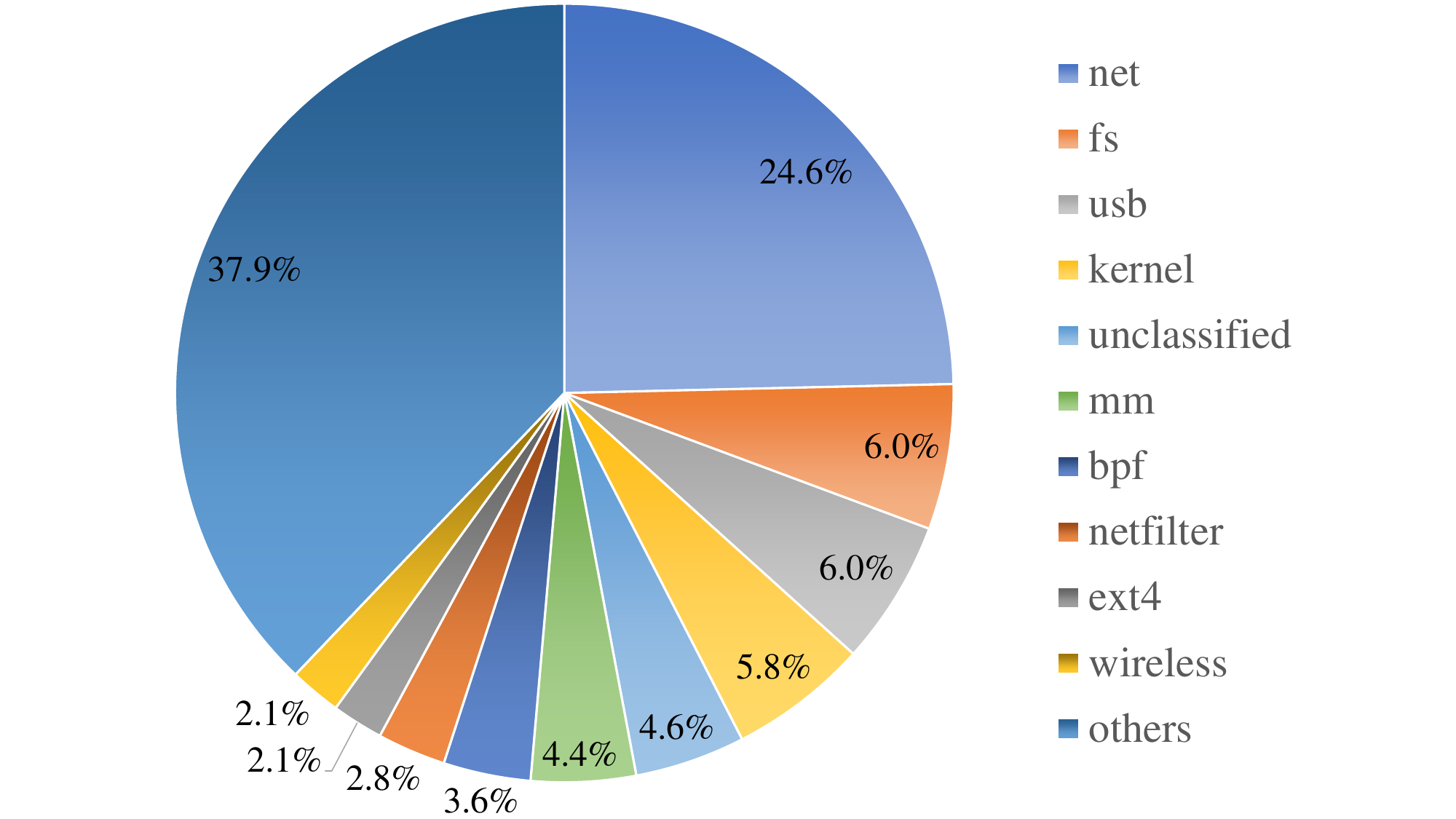}
\end{center}
\vspace{-0.3em}
\caption{\label{fig:vul} Distribution of Linux kernel vulnerabilities reported by Syzkaller~\cite{syzkaller}. Only the top 10 subsystems with the most vulnerabilities are listed for demonstration.}
\vspace{-1.5em}
\end{figure}

A more critical problem is that most previous works neglect the support for bi-directional isolation~\cite{minibox14bi, sgxlock22} within the kernel space. They have to trust the core kernel for management and only enforce one-way access control between the core kernel and other untrusted components. The untrusted parts are restricted from accessing resources of the core kernel, while in the opposite direction, the trusted core kernel can arbitrarily access other components, leading to the monopoly over isolation~\cite{hotos23trust}. Unfortunately, our analysis revealed that about 5.8\% of Linux vulnerabilities are found in the core kernel (\autoref{fig:vul}), which is comparable to notorious subsystems like \cc{ipv6} and \cc{netfilter}.

In addition to direct exploitation in the core kernel, one-directional isolation leaves confused deputy attacks as the open problem~\cite{Lefeuvre2023, civscope23, hotos23interface}. Since the trusted part is unhindered, it can be confused by malicious inputs to break compartmentalization on behalf of the untrusted compartment. Therefore, both the source and the target of cross-compartment interactions should be validated bi-directionally.

This paper introduces \sys, a comprehensive framework for kernel compartmentalization that enforces bi-directional isolation with minimal performance overhead. \sys leverages Intel's recent hardware feature, Protection Keys for Supervisor-mode (PKS)~\cite{intel23manual}, in a novel way to isolate data and code into mutually untrusted compartments. We exclude the core kernel from the TCB and tag it as well as other compartments with distinct protection keys (\cc{pkey}s). A newly introduced lightweight in-kernel monitor is in charge of enforcing security invariants, including data integrity, execute-only memory (XOM), and compartment interface integrity that defends against confused deputy attacks according to developer-specified policies flexibly. With a specially designed switch gate table (SGT) and zero-copy ownership transfer, the compartment switching is performed securely and efficiently.

Employing PKS in kernel compartmentalization presents unique challenges due to the privileged environment. First, the memory access validation of PKS does not work on execution permissions, leaving the control flow not validated and unprotected~\cite{iskios21}. Second, since all kernel code shares the privilege of modifying page tables (PTs) and the \cc{PKRS} register, PKS-based isolation suffers from pitfalls~\cite{pitfall20, voulimeneas2022you, schrammel2022jenny} to be bypassed through PT tampering or instruction abusing. Third, the 4-bit \cc{pkey} supports only up to 16 compartments, which is inadequate. The Linux kernel v6.1 has over 30 million lines of code and contains 6296 LKMs with the x86\_64 generic configuration. The limited number of compartments makes each compartment too coarse to confine the impact of vulnerability exploitation. 

To address these challenges, on the security hand, \sys introduces a lightweight in-kernel monitor through privilege separation for invariant enforcement. It restricts all kernel code pages as XOM, enables diversification schemes such as Kernel Address Space Layout Randomization (KASLR)~\cite{KASLR}, isolates private stacks, and enforces cross-compartment control flow integrity (CFI). These together substantially mitigate control flow hijacking. The page table and switch gates are exclusively controlled by the monitor to prevent PKS-based isolation from being bypassed. On the scalability hand, \sys proposes a locality-aware 
two-level scheme to support unlimited compartments. Different from the two-level compartmentalization in HAKC~\cite{mckee2022preventing}, which combines the ARM-specific features: Memory Tagging Extension (MTE) and Pointer Authentication (PA), we group the modules and split the address space based on the locality of module interactions. The first level is PKS-based intra-address space isolation, while the second level is locality-aware inter-address space isolation with Address Space Identifier (ASID). 

We implement a prototype system in Linux v6.1 and perform automated LKM isolation as a use case. In the case study, we specify security policies based on the boundary analysis, shared data analysis, and security check analysis, then, compartmentalize all 160 LKMs with \cc{localmodconfig} supported by our machine to show the scalability. Security analysis with penetration tests confirms its comprehensive protection capabilities. Extensive performance evaluation with both micro-benchmarks and macro-benchmarks presents its efficiency. Specifically, \sys incurs an average overhead of only 2.44\% on the whole-system benchmarks Phoronix~\cite{phoronix} and ApacheBench~\cite{ab} tests on the compartmentalized \cc{ipv6} module show an overhead of less than 2\%.

In summary, this paper makes the following contributions:
\begin{itemize}
    \item A systematic analysis of kernel compartmentalization objectives and a secure, scalable, and efficient mechanism based on PKS that fulfills these objectives.
    \item A novel in-kernel monitor that supports bi-directional isolation and enforces multiple security-critical invariants, including data integrity, execute-only memory, and compartment interface integrity.
    \item A locality-aware two-level compartmentalization scheme that supports unlimited compartments.
    \item An implementation of the prototype system\footnote{Available at \url{https://github.com/gyg128/BULKHEAD}} for automated LKM isolation and extensive evaluation that demonstrates its security, scalability, and efficiency.
\end{itemize}

\section{Objectives For Kernel Compartmentalization}
\label{sec-obj}

\begin{table*}[htbp]
\footnotesize
\begin{center}
\resizebox{\textwidth}{!}{%
\begin{tabular}{@{}
>{\columncolor[HTML]{FFFFFF}}l |
>{\columncolor[HTML]{FFFFFF}}l |
>{\columncolor[HTML]{FFFFFF}}l 
>{\columncolor[HTML]{FFFFFF}}l 
>{\columncolor[HTML]{FFFFFF}}l 
>{\columncolor[HTML]{FFFFFF}}l |
>{\columncolor[HTML]{FFFFFF}}l |
>{\columncolor[HTML]{FFFFFF}}l 
>{\columncolor[HTML]{FFFFFF}}l |
>{\columncolor[HTML]{FFFFFF}}l @{}}
\toprule
\cellcolor[HTML]{FFFFFF} &
  \multicolumn{1}{c|}{\cellcolor[HTML]{FFFFFF}} &
  \multicolumn{4}{c|}{\cellcolor[HTML]{FFFFFF}\textbf{Security}} &
  \multicolumn{1}{c|}{\cellcolor[HTML]{FFFFFF}\textbf{Scalability}} &
  \multicolumn{2}{c|}{\cellcolor[HTML]{FFFFFF}\textbf{Performance}} &
  \multicolumn{1}{c}{\cellcolor[HTML]{FFFFFF}} \\ \cmidrule(lr){3-9}
\multirow{-2}{*}{\cellcolor[HTML]{FFFFFF}} &
  \multicolumn{1}{c|}{\multirow{-2}{*}{\cellcolor[HTML]{FFFFFF}\textbf{Mechanisms}}} &
  \begin{tabular}[c]{@{}l@{}}bi-directional\\ isolation\end{tabular} &
  \begin{tabular}[c]{@{}l@{}}data\\ protection\end{tabular} &
  \begin{tabular}[c]{@{}l@{}}control flow\\ protection\end{tabular} &
  \begin{tabular}[c]{@{}l@{}}interface\\ protection\end{tabular} &
  \begin{tabular}[c]{@{}l@{}}domain\\ number\end{tabular} &
  \begin{tabular}[c]{@{}l@{}}domain\\ switch\end{tabular} &
  \begin{tabular}[c]{@{}l@{}}data\\ transfer\end{tabular} &
  \multicolumn{1}{c}{\multirow{-2}{*}{\cellcolor[HTML]{FFFFFF}\textbf{Compatibility}}} \\ \midrule
seL4~\cite{klein2009sel4} &
  Microkernel &
  No &
  Yes &
  Yes &
  No &
  Unlimited &
  Low &
  Low &
  Heavy redesign \\
UnderBridge~\cite{gu2020harmonizing} &
  Microkernel+PKU &
  No &
  Yes &
  Yes &
  No &
  16 &
  High &
  High &
  Heavy redesign \\
LXFI~\cite{mao2011software} &
  SFI &
  No &
  Yes &
  Yes &
  Yes &
  Unlimited &
  Low &
  Low &
  Annotations \\
LVD~\cite{lvd20} &
  Virtualization &
  No &
  Yes &
  Yes &
  No &
  512 &
  High &
  Low &
  Nested Virtualization \\
KSplit~\cite{huang2022ksplit} &
  Virtualization &
  No &
  Yes &
  Yes &
  No &
  512 &
  High &
  Low &
  Nested Virtualization \\
xMP~\cite{proskurin2020xmp} &
  Virtualization &
  No &
  Yes &
  No &
  No &
  512 &
  High &
  Low &
  Nested Virtualization \\
Nested Kernel~\cite{dautenhahn2015nested} &
  WP bit &
  No &
  Yes &
  Yes &
  No &
  2 &
  High &
  High &
  x86-64 \\
SKEE~\cite{azab2016skee} &
  PT switching&
  No &
  Yes &
  Yes &
  No &
  2 &
  Medium &
  Low &
  ARM \\
IskiOS~\cite{iskios21} &
  PKU &
  No &
  No &
  Yes &
  No &
  8 &
  High &
  High &
  SMAP/SMEP \\
HAKC~\cite{mckee2022preventing} &
  MTE+PA &
  No &
  Yes &
  Yes &
  No &
  Unlimited &
  Medium &
  Medium &
  ARM \\
CHERI~\cite{watson2015cheri} &
  New architecture &
  No &
  Yes &
  Yes &
  Yes &
  Unlimited &
  Medium &
  Medium &
  New architecture \\
SecureCells~\cite{bhattacharyya2023securecells} &
  New architecture &
  No &
  Yes &
  Yes &
  Yes &
  Unlimited &
  High &
  High &
  New architecture \\
DOPE~\cite{dope23} &
  PKS &
  No &
  Yes &
  No &
  No &
  16 &
  High &
  High &
  Intel \\
\textbf{BULKHEAD} &
  \textbf{PKS} &
  \textbf{Yes} &
  \textbf{Yes} &
  \textbf{Yes} &
  \textbf{Yes} &
  \textbf{unlimited} &
  \textbf{High} &
  \textbf{High} &
  \textbf{Intel} \\ \bottomrule
\end{tabular}%
}
\end{center}
\caption{Systematic analysis of kernel compartmentalization objectives.}
\label{tb:objectives}
\vspace{-1.5em}
\end{table*}

Kernel compartmentalization mechanisms should be characterized by a specific set of objectives, including security, scalability, performance, and compatibility issues. With a systematic survey (\autoref{tb:objectives}), we analyze the limitations of related work and demonstrate how \sys meets these goals.

\subsection{Kernel Vulnerability Analysis}
\label{sec2-vul}
As inspiration for \sys, we analyzed all fixed vulnerabilities reported by Syzkaller~\cite{syzkaller} - one of the most popular kernel fuzzing tools developed by Google, with 5201 in total at the time of writing. According to the location of the patches, we investigate the distribution of vulnerabilities, as shown in \autoref{fig:vul}. For ease of demonstration, only the top 10 subsystems with the highest number of vulnerabilities are listed. As we can see, there are about 5.8\% of Linux vulnerabilities found in the core kernel, which ranks fourth. Specifically, the number of issues from the core kernel is comparable to some notorious subsystems such as \cc{ipv6} and \cc{netfilter}. Given that any single vulnerability has the potential to destroy the entire system, simply trusting the core kernel is highly questionable. 

\subsection{Security Objectives}
\label{sec-obj1}
To realize the security goal, a mechanism must enforce comprehensive protection of the isolated compartments.

\subsubsection{Bi-directional Isolation} 
Despite the significant risks posed by vulnerabilities within the core kernel, most existing efforts in kernel compartmentalization~\cite{lvd20, mckee2022preventing, huang2022ksplit} only perform one-directional isolation at the boundaries between the core kernel and other compartments. They grant arbitrary access to the core kernel and only restrict other compartments due to the challenge of intra-kernel privilege separation~\cite{dautenhahn2015nested}. The reverse protection is hard because these approaches have to rely on the core kernel to manage and enforce access control policies. Besides direct vulnerability exploitation in the core kernel, the trusted part can also be exploited as a confused deputy, thereby necessitating a more robust strategy under the mutually untrusted threat model. Effective compartmentalization must ensure bi-directional isolation among all compartments and rigorously validate both the source and the target for each cross-compartment interaction. Several works have explored bi-directional protection at kernel-userspace boundaries~\cite{minibox14bi}, hypervisor-VM boundaries~\cite{hypsec19, cloud20}, and application-enclave boundaries~\cite{sgxlock22, biIso23}. While we enforce PKS-based bi-directional isolation within the kernel space, which breaks the monopoly of the core kernel without introducing any additional layer.

\subsubsection{Data Protection}
Sensitive data objects in the kernel, especially privilege-related objects like \cc{cred}, are critical attack targets~\cite{lin2022dirtycred}. Kernel compartmentalization must isolate data into specific compartments and block any illegal access from irrelevant compartments. Existing works, such as IskiOS~\cite{iskios21} and KCoFI~\cite{kcofi14}, leave sensitive data protection out of consideration and only focus on control flow protection. However, data-oriented attacks could also indirectly change the control flow and even escalate the privilege~\cite{dop16}. \sys assigns different \cc{pkey}s to objects of different compartments, which guarantees the data integrity of each compartment with hardware-enforced access control.

\subsubsection{Control Flow Protection}
Control flow hijacking is another major threat to kernel security. While demanding validity checks for control flow transfer, many kernel isolation works, such as xMP~\cite{proskurin2020xmp} and DOPE~\cite{dope23}, rely on additional control flow protection mechanisms as assumptions, such as CFI~\cite{kcofi14}. Unfortunately, a simple combination of data protection and control flow protection will lead to significant performance overhead. As a countermeasure, \sys set all kernel code regions as XOM to mitigate code reuse attacks within compartments. While cross-compartment calls are guarded by carefully designed switch gates. Although XOM itself is not a strong defense, existing diversification schemes such as KASLR and data compartmentalization make memory disclosure~\cite{disclosure16} harder to break our protection. 

\subsubsection{Compartment Interface Protection}
Compartment interface security is a long-neglected problem. Recent work~\cite{Lefeuvre2023} has emphasized the seriousness of interface-related vulnerabilities, which could be used to launch confused deputy attacks. With malicious inputs, attackers aim to indirectly exploit the privilege of the confused callee. Most existing works cannot defend against confused deputy attacks with one-directional isolation. Weak interfaces seriously reduce or even fully negate the security guarantee of compartmentalization. LXFI~\cite{mao2011software} proposed API integrity to protect kernel API, but it relies on the programmer's annotations, which require a lot of manual effort and are error-prone. CHERI~\cite{watson2015cheri} and SecureCells~\cite{bhattacharyya2023securecells} also discussed interface checks. However, as new hardware architectures, it is still far from actual adoption in practice. \sys introduces compartment interface integrity with the help of developer-specified policies. With the novel switch gate table (SGT) protected by the in-kernel monitor, we record and validate both the source and the target compartment metadata. This bi-directional validation grants \sys the ability to protect compartment interfaces according to the policies specified by developers. For example, we define entry/exit points for compartment switches through boundary analysis and add validations for data transfer based on shared data and security check analysis.

\subsection{Scalability Objectives}
\label{sec-obj2}
\subsubsection{Scalable for Unlimited Compartments}
The security benefits of compartmentalization depend on the granularity of the compartments, which is limited by the number of supported domains. Nested Kernel~\cite{dautenhahn2015nested} only constructs two domains based on the Write Protection (WP) bit. SKEE~\cite{azab2016skee} creates a separate secure execution environment within the kernel through PT switching. Memory Protection Key (MPK)-based approaches suffer from the hardware limitation of 16 \cc{pkey}s. Reserving \cc{pkey} for other purposes will further minimize the domains available to the kernel~\cite{iskios21}. The virtualization-based solutions~\cite{lvd20, proskurin2020xmp, huang2022ksplit} support up to 512 EPTs representing 512 different memory views. HAKC~\cite{mckee2022preventing} proposed a two-level compartmentalization scheme combining the ARM-specific features MTE and PA. Although it theoretically supports unlimited compartments, HAKC evaluated only two compartments and got a linear growth of 14\%-19\% per compartment in overhead, attributed to the expensive cryptographic operations of PA. As a comparison, \sys proposes a locality-aware two-level compartmentalization scheme combining PKS and multiple address spaces. In addition to PKS-based intra-address space isolation, we also leverage locality-aware inter-address space isolation with ASID. This second level is more compatible and efficient than HAKC. Since each address space has its own 16 compartments, we can achieve support for unlimited compartments by address space switching.

\subsection{Performance Objectives}
\label{sec-obj3}
\subsubsection{Fast Compartment Switches}
Since the kernel serves a central role in the system, it is extremely sensitive to performance overhead. Compartment switches are essential and frequent for interactions, which dominate the performance. Microkernels~\cite{klein2009sel4, gu2020harmonizing} suffer from time-consuming IPC. The inserted security checks for every memory access slow down the performance of SFI-based approaches~\cite{mao2011software}. Although the \cc{vmfunc} instruction accelerates EPT switching~\cite{lvd20, proskurin2020xmp, huang2022ksplit}, due to nested paging and I/O virtualization, running systems in VMs additionally incurs extra overhead~\cite{vmPerformance15, performance21}. \sys benefits from efficient permission validations and switches with PKS. The hardware-based compartment switch only involves a specific register update.

\subsubsection{Zero-copy Data Transfer}
Data copying across compartments can overwhelm the performance-critical OS kernel. Unfortunately, PT/EPT switching-based approaches cannot support secure zero-copy communication across different memory mapping and require complex synchronization of shared states~\cite{azab2016skee, lvd20, huang2022ksplit}. In contrast, tag-based approaches inherently support zero-copy data transfer. \sys tags data with a specific \cc{pkey}, which enforces single ownership of objects. As a response to compartment switching, access from the target compartment will trigger a page fault. Then, the monitor validates the shared data and just updates its \cc{pkey} in the dedicated handler, without copying data across compartments.

\subsection{Compatibility Objectives}
\label{sec-obj4}
\subsubsection{Compatibility Issues}
Besides security, scalability, and performance, ideal kernel compartmentalization mechanisms also should be compatible with real-world machines and complex production environments. Different from related work, \sys utilizes the emerging hardware feature of the widely-used Intel architecture. It requires neither heavy code redesign like the microkernel~\cite{klein2009sel4, gu2020harmonizing} nor extensive annotations like LXFI~\cite{mao2011software}. Moreover, the compartmentalized kernel based on PKS could be directly used in the cloud environment without nested virtualization restrictions~\cite{lvd20, huang2022ksplit, proskurin2020xmp}.

\section{Background}
\label{sec2}


\subsection{Memory Protection Keys}
\label{sec2-mpk}

\begin{figure}
\begin{center}
\includegraphics[width=.95\linewidth]{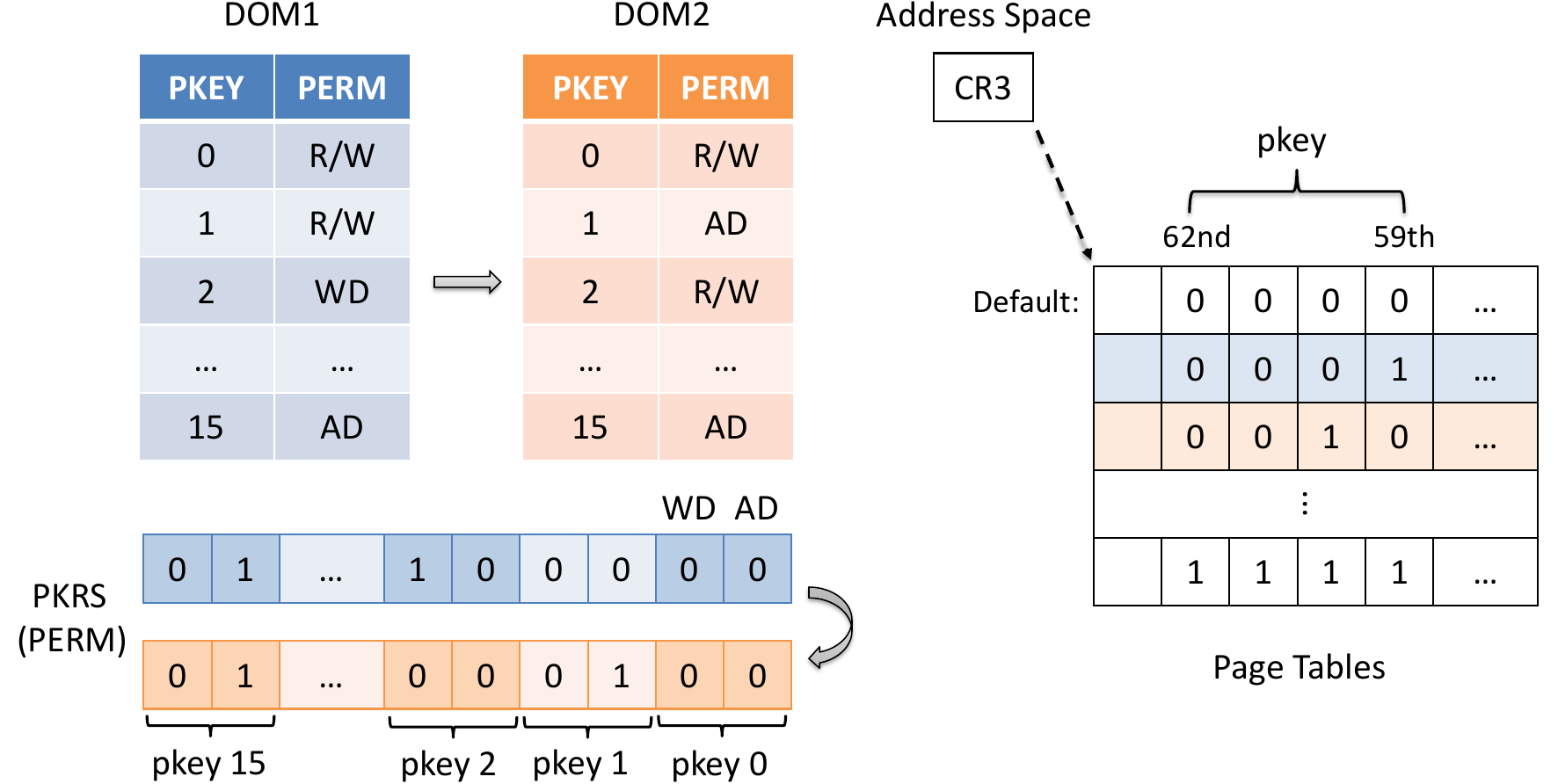}
\end{center}
\caption{\label{fig:mpk} Working principle of MPK, where WD and AD stand for write-disable and access-disable permissions, respectively.}
\vspace{-1.5em}
\end{figure}

MPK~\cite{intel23manual} is an Intel hardware feature that enforces per-thread access control but without requiring PT modification for domain switching (\autoref{fig:mpk}). In addition to the permission bits (R/W bit and NX bit) in the page table entries (PTE), it tags pages with a 4-bit \cc{pkey} (bits [59:62]), which partitions the address space into at most 16 memory domains. The permissions of pages with a specific \cc{pkey} are stored in a dedicated per-thread register as a 2-bit notation (WD, AD), where WD means write-disable and AD means access-disable. The hardware-based access checks incur nearly zero runtime overhead~\cite{gu2020harmonizing}, while permission change is simply done by updating the register. Since no PT walk or TLB flush is required, MPK is known for the fast domain switching~\cite{vahldiek2019erim, hedayati2019hodor}.

According to the User/Supervisor (U/S) bit in the PTE, MPK has two variants, namely Protection Keys for User-mode (PKU) and Protection Keys for Supervisor-mode (PKS), using protection key rights register for user pages (\cc{PKRU}) and protection key rights register for supervisor pages (\cc{PKRS}, i.e., \cc{MSR 0x6E1}) respectively. PKU has been explored in depth for intra-process isolation and has shown outstanding results~\cite{hedayati2019hodor, vahldiek2019erim}. Before PKS was available, several works attempted to utilize PKU for kernel isolation by setting the User bit of kernel pages, such as UnderBridge~\cite{gu2020harmonizing} and IskiOS~\cite{iskios21}. This unconventional application requires additional consideration for kernel-user isolation and is incompatible with other security features based on the U/S bit, such as Supervisor Mode Access Prevention (SMAP) and Supervisor Mode Execution Prevention (SMEP). Multiple untrusted compartments and the privileged environment also pose challenges to PKU-based kernel isolation, especially the switch gate design.

The long-awaited feature PKS has recently become available on 12th Core and 4th Xeon CPUs~\cite{intel23manual}, yet it remains underexplored in existing works. There are some PKS-based studies concurrent to \sys. Among them, KDPM~\cite{kuzuno2022kdpm} and DOPE~\cite{dope23} only target kernel data protection, neglecting other security objectives. MOAT~\cite{moat24} and eBPF-sandbox~\cite{pksjos} are specific to eBPF program isolation and cannot generalized to other kernel components. To the best of our knowledge, \sys is the first work that achieves comprehensive kernel compartmentalization with PKS, addressing both security and scalability challenges systematically as detailed in \autoref{sec-design}.

\section{Threat Model}
\label{sec3}

We assume that vulnerabilities are present throughout the kernel, encompassing both the core kernel and LKMs. A non-root adversary can exploit these to launch various attacks, such as data-oriented attacks, control flow hijacking, and confused deputy attacks. The ultimate goal of this potential attacker is to spread the impact of vulnerabilities and finally take control of the entire system. Consequently, kernel components are mutually untrusted and the core kernel is also excluded from the TCB. We trust the lightweight in-kernel monitor for security invariant enforcement, and its correctness can be formally verified. We trust the system developer who specifies prior compartmentalization policies. However, users who load compartments and register gates at runtime could be malicious and their behaviors are validated by the monitor against the predefined policies. We also assume the secure boot mechanism~\cite{suh2003aegis, wilkins2013uefi} and trust the underlying hardware.

We focus on software attacks and do not consider physical attacks such as cold boot attacks~\cite{yitbarek2017cold} and RawHammer attacks~\cite{mutlu2019rowhammer}. Denial-of-service (DoS) and side-channel attacks like Spectre~\cite{spectre19} and Meltdown~\cite{meltdown18} are also excluded from our consideration. The defense mechanisms against these attacks are orthogonal to kernel compartmentalization and can be applied to protect the system further.

\section{\sys Design} 
\label{sec-design}

\subsection{Overview}
\label{sec-overview}

\begin{figure}
\begin{center}
\includegraphics[width=\linewidth]{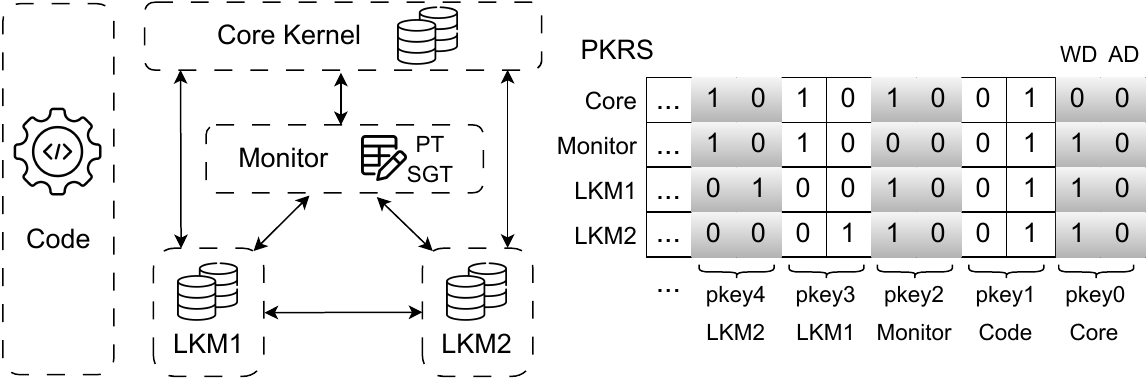}
\end{center}
\caption{\label{fig:overview} The overview of \sys.}
\vspace{-1.5em}
\end{figure}

\sys provides strong protection by adhering to the principle of least privilege. As shown in \autoref{fig:overview}, the kernel memory is partitioned and tagged with different \cc{pkey}s, and the per-thread \cc{PKRS} indicates the access permissions for each compartment. Cross-compartment access is controlled bi-directionally, thus all compartments, including the core kernel and the monitor are restricted to accessing only what is necessary for their operations~(\autoref{sec-bi}). The data of each compartment can only be modified by itself. All kernel code pages are tagged as XOM. It is hard to reuse gadgets without a readable code layout. While the lightweight in-kernel monitor~(\autoref{sec-monitor}) serves as a special compartment for security invariant enforcement~(\autoref{sec-invariant}). Specifically, the page table (PT) and a novel switch gate table (SGT) are write-protected by the monitor, and compartment switches are performed securely and efficiently based on the SGT. Although we take two LKMs as an example in \autoref{fig:overview}, \sys incorporates a locality-aware two-level compartmentalization scheme to support unlimited compartments~(\autoref{sec-scalability}). 

\subsection{PKS-based Bi-directional Isolation}
\label{sec-bi}

Bi-directional isolation handles mutual distrust through cross-compartment access control. Each compartment is restricted to accessing only necessary resources, even for the core kernel and the monitor. As \autoref{fig:overview} shows, the memory of each compartment is tagged with a distinct \cc{pkey}. All kernel code pages are attached with \cc{pkey1} to realize XOM. The default \cc{pkey0} represents the core kernel’s data, while \cc{pkey2} is assigned to the monitor. The PT and SGT are exclusively owned by the monitor and read-only for other compartments. \sys configures the (WD, AD) notations in the per-thread \cc{PKRS} for bi-directional access control. The \cc{PKRS} value uniquely identifies the current compartment by only one (0,~0) notation which means full access to memory with the respective \cc{pkey}. Notations of other \cc{pkey}s all have the WD or AD bits set to enforce the compartmentalization policy. As a result, compartments, including the core kernel, are mutually untrusted, and no one can arbitrarily access each other's memory. Any attempt to break the access permissions will result in a protection key violation fault. Compartment switches can only be performed by switch gates registered in the SGT that alter the \cc{PKRS} to disable access to the source compartment and grant access to the target compartment. Both the source and the target are validated according to the metadata recorded in the write-protected SGT, thus mitigating confused deputy attacks.

\subsection{In-kernel Monitor}
\label{sec-monitor}
The lightweight monitor is a special compartment responsible for guaranteeing bi-directional isolation. It manages critical metadata such as the PT for permission restriction and the SGT for secure switching. Unlike virtualization-based monitors\cite{lvd20, huang2022ksplit}, \sys does not rely on an additional layer and thus breaks the "turtles all the way down" paradigm. The monitor is constructed through privilege separation within the same level of the kernel. We protect the memory resources of the monitor by PKS-based isolation while securing the instruction and register resources by depriving other compartments' privileges.

\subsubsection{Memory Isolation for In-Kernel Monitor}
With the help of PKS, the monitor memory is isolated from the rest of the kernel. To be specific, the data part especially the PT and the SGT is tagged with \cc{pkey2}, thus write-protected against other compartments by the WD bit in \cc{PKRS}. Only the monitor has the privilege to update these metadata related to access control. On the other hand, the code part is tagged with \cc{pkey1} as XOM so that attackers cannot reuse the privileged code maliciously. Note that although the monitor is trusted, it also cannot access other compartments directly due to bi-directional isolation.

\subsubsection{Instruction Deprivation}
Besides memory resources, instruction and register resources can also be abused to break the compartmentalization~\cite{fan2023isa}, such as malicious \cc{PKRS} updates. Moreover, since x86 does not require instruction alignment, the unintentional occurrences of privileged instructions, such as part of a longer instruction or spanning two consecutive instructions, are also hazardous. \sys should restrict other compartments' behaviors that run with the monitor in the same privilege level (Ring-0) by instruction deprivation. 

Benefiting new advances in binary rewriting~\cite{vahldiek2019erim, wu2022dancing}, we eliminate privileged instructions in compartments other than the monitor. At a high level, unintended occurrences are replaced with functionally equivalent instructions, while intended occurrences are replaced with switch gates to the monitor. As a result, only the lightweight monitor has the capability to execute the security-critical instructions after validation. Even if attackers attempt to reuse these occurrences in the monitor, they cannot locate the useful gadgets because of the read-disable protection of XOM. Specifically, we focus on three categories of instructions in our prototype. First, the instruction that changes the value of \cc{PKRS}, i.e., \cc{wrmsr}. Second, instructions that write control registers, i.e., \cc{mov-to-CRn}. For example, attackers can forge the page table with a malicious value of \cc{CR3} or disable PKS by clearing the PKS bit in \cc{CR4}. Third, instructions that store system registers like \cc{IDTR}, \cc{GDTR}, \cc{LDTR}, etc. It is notable that malicious interrupts are defended by the atomic switch gate described in \autoref{I3} instead of eliminating the 1-byte interrupt enabling/disabling instruction. 

\begin{figure}
\begin{center}
\includegraphics[width=.85\linewidth]{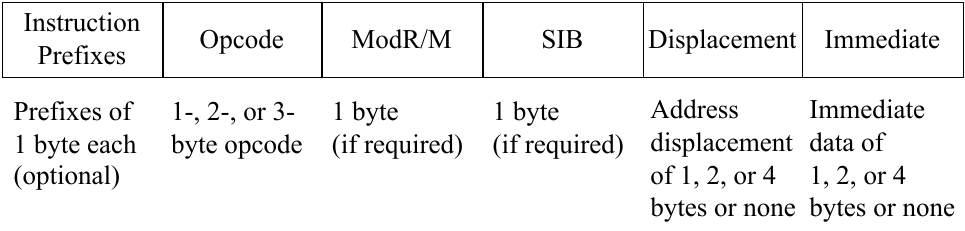}
\end{center}
\vspace{-0.6em}
\caption{\label{fig:instruction} Intel 64 and IA-32 architectures instruction format.}
\vspace{-0.6em}
\end{figure}

\begin{figure}
\begin{center}
\includegraphics[width=\linewidth]{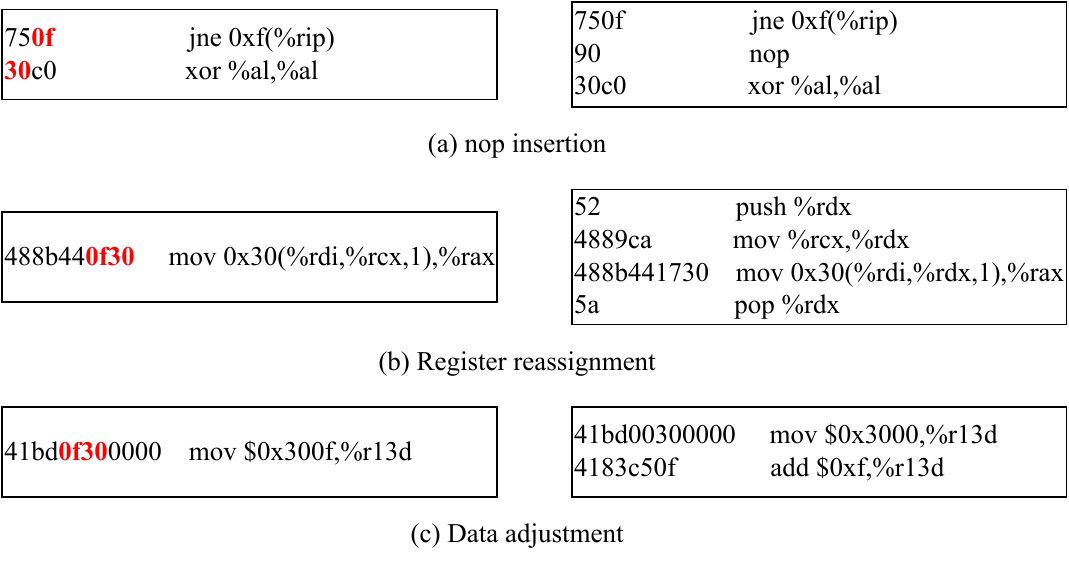}
\end{center}
\vspace{-0.5em}
\caption{\label{fig:rewriting} Some examples of eliminating unintended \cc{wrmsr} (\cc{0x0f30}).}
\vspace{-1.2em}
\end{figure}

For unintended instructions, we apply different binary rewriting strategies according to the fields with which the unintended subsequence overlaps. The Intel 64 and IA-32 architectures instruction encodings are subsets of the format shown in \autoref{fig:instruction}~\cite{intel23manual}. An instruction consists of: (1) an optional instruction prefix, (2) a primary opcode field (up to three bytes), (3) a ModR/M field that determines the addressing form and includes a register operand, (4) a SIB (Scale-Index-Base) field that specifies registers for indirect memory addressing, (5) a displacement and/or (6) an immediate data field that specify constant offsets. If an unintended sequence spans two or more instructions, it could be broken by inserting a 1-byte \cc{nop} (\cc{0x90}) or reordering instructions. Otherwise, unintended instructions may appear entirely within a longer instruction. For these cases, we eliminate them by register reassignment, data adjustment, or replacement with other functionally equivalent instructions. These strategies are applied iteratively until there are no unintended instructions. \autoref{fig:rewriting} shows some concrete examples of eliminating unintended occurrences of \cc{wrmsr} (\cc{0x0f30}).

For intended instructions, we replace each occurrence with a switch gate to the monitor and delegate the monitor to execute the instruction after security validation. The validation of the switch gate relies on developer-specified policies and is flexible to perform various checks. In particular, we record rich metadata on the source and the target as described in \autoref{I3}. Thus, except for the lightweight monitor protected by XOM, other domains are deprived of the capability to bypass isolation through privileged instructions. Although \sys could benefit from hardware extensions~\cite{fan2023isa} for instruction deprivation, binary rewriting offers compatibility with existing hardware.

\subsection{Invariant Enforcement}
\label{sec-invariant}
With intra-kernel privilege separation, the monitor enforces a series of invariants to fulfill the security objectives: (1)~data integrity that defends data-oriented attacks, (2) execute-only memory that mitigates control flow hijacking, and (3) compartment interface integrity targets confused deputy attacks.

\subsubsection{Data Integrity}
\label{I1}
\textit{Data of each compartment can only be modified by itself. \textbf{\textit{(I1)}}}

\sys determines the ownership of data according to the \cc{pkey}. By assigning different \cc{pkey}s to different compartments in PTEs, the monitor guarantees no compartments share the same \cc{pkey}. Since the PTEs contain \cc{pkey}s as well as other permission control bits, the page table is a natural prime target of attackers. To prevent arbitrary tampering, the page table is read-only for other compartments. All PT updates are exclusively delegated to the monitor, and only those adhering to the compartmentalization policy can be performed. With instruction deprivation, attackers also cannot forge page tables by malicious \cc{CR3} value. As a result, \sys enforces PT integrity to support data integrity: \textit{the page table can only be modified by the monitor. \textbf{\textit{(I1.1)}}} \label{I1.1}

\sys comprehensively protects the compartment's memory, including heap, stack, physmap, and MMIO regions. For the objects allocated dynamically at runtime, we create a private heap for each compartment. An intuitive way is to modify the \cc{pkey} in the PTE after allocation. However, this dynamic updating through page walk is expensive. Instead, during the initialization of a specific compartment, we reserve a tagged memory pool for it. Then, the allocators are modified to use the reserved cache without the overhead of dynamic tagging. With the private heap, cross-compartment heap corruption becomes impossible. Similarly, we also provide private stacks for compartments and switch the stack during compartment switching. As a result, attackers cannot disturb other compartments' stacks to hijack the control flow. Besides, the direct map of physical memory (physmap) and the memory-mapped I/O (MMIO) regions are also tagged with \cc{pkey}. Access to these areas goes through PKS hardware validation as well.

\subsubsection{eXecute-Only Memory (XOM)}
\label{I2}
\textit{All kernel code cannot be read or written by anyone. \textbf{\textit{(I2)}}}

XOM is a lightweight but effective control flow protection~\cite{kernelXOM19}. Although PKS does not perform hardware-enforced checks on execution permission, we set the AD bit in \cc{PKRS} for all kernel code regions to disable any write or read access, which is required by code injection or code reusing.

With the $W\bigoplus X$ policy in the kernel, all executable regions are not writable, which prevents code injection attacks. While code reuse attacks (e.g., ROP~\cite{rop12}) highly rely on information about where and how code has been placed in memory. Combined with diversification schemes like KASLR, read-disabling hinders the attacker from finding useful gadgets. As a special case, a few code pages need dynamic updates at runtime. For example, the code generated by the BPF Just-In-Time (JIT) compiler. These update requests are also directed to the monitor. After validation, \sys will temporarily grant the monitor access to these pages. Albeit whole kernel CFI provides stronger security, it comes at the cost of performance~\cite{kcofi14, finecfi18}. Besides, constructing a precise global control flow graph (CFG) for the huge kernel is still challenging~\cite{mlta19lu, typro22}. As a trade-off, we enforce compartment interface integrity for the cross-compartment call. Cross-compartment CFI and private stack make XOM offer better security than itself.

\subsubsection{Compartment Interface Integrity (CII)}
\label{I3}
\textit{Compartment switches must occur at the predefined entry/exit points and pass data according to security policies. \textbf{\textit{(I3)}}}

Previous work~\cite{Lefeuvre2023, civscope23, hotos23interface} has assessed the serious impact of compartment interface vulnerabilities (CIVs), where a compartment can be exploited as a confused deputy to control the execution or corrupt data of other compartments. Accordingly, we propose CII against these threats on the basis of bi-directional isolation. First, the control flow is constrained by the predefined entry/exit points. Second, the shared data is validated according to security policies.

\begin{figure}
\begin{center}
\includegraphics[width=.92\linewidth]{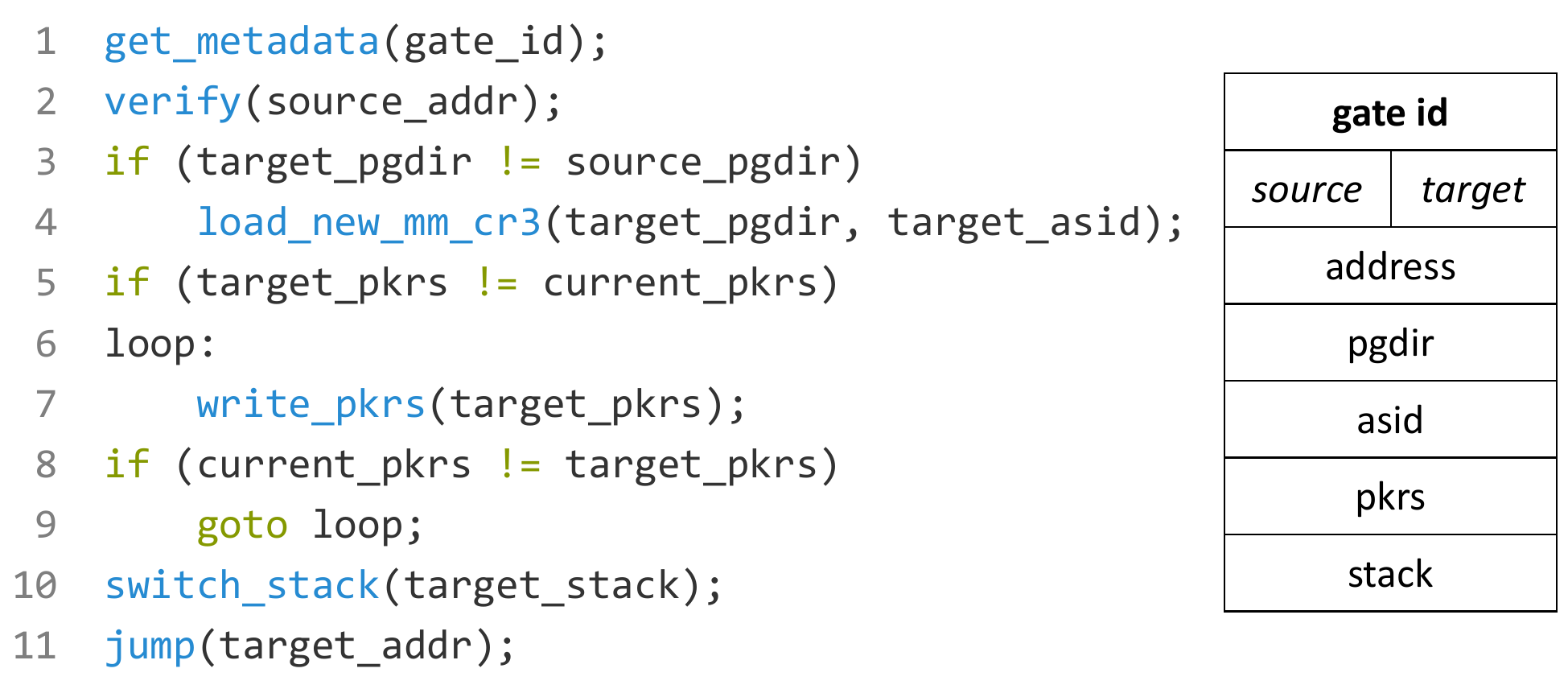}
\end{center}
\caption{\label{fig:switch} The switch gate pseudocode (left) and the metadata format of a SGT entry (right).}
\vspace{-1.5em}
\end{figure}

\PP{Secure Switching}
\sys's monitor maintains a software SGT inspired by some hardware extensions for secure domain switching~\cite{fan2023isa, bhattacharyya2023securecells}. Different from other user-mode switch gate designs, such as ERIM~\cite{vahldiek2019erim} and Hodor~\cite{hedayati2019hodor}, the unique challenges of kernel compartmentalization come from multiple untrusted compartments and the privileged environment. As shown in \autoref{fig:switch}, the novel SGT contains informative metadata of all registered gates and directs each switch to the target address of the legal compartment. Secure switching implies that all switch gates satisfy the properties of atomicity, determinism, and exclusivity, which prevent malicious exploitation of interfaces and thus support CII.

\textbf{\textit{P1.}} \textit{Atomicity: Compartment switches through switch gates are executed atomically.} \label{P1} 

Atomicity means that the series of operations shown in \autoref{fig:switch} cannot be exploited separately. The switch gate first uses a registered gate id to retrieve the gate metadata from the write-protected SGT. According to the metadata, it verifies the source address and updates \cc{CR3} if the source and the target compartment are in different address spaces (i.e., two-level compartmentalization described in \autoref{sec-scalability}). Then, it changes \cc{PKRS} to the target value on demand by \cc{wrmsr}, switches to the private stack of the target compartment, and finally jumps to the target address recorded in the SGT. Here the most crucial step is the \cc{PKRS} update, which implies the permission transfer. 

Attackers can hijack the execution of the switch gate in two ways: (1) interrupting the sequence, (2) jumping to the middle of the sequence. First, malicious interrupts after \cc{PKRS} updates can make the attacker gain the privilege of the original target compartment. To prevent this, we update \cc{PKRS} to the default restricted value as the core kernel's privilege at each interrupt entry so that potentially illegal \cc{PKRS} will not work for the interrupt context. Second, since the value of \cc{PKRS} is not updated directly by an immediate operand but taken from the \cc{eax} register, attackers may jump to just before the \cc{wrmsr} instruction with forged \cc{eax} value. As a countermeasure, we add a security check after \cc{wrmsr} to guarantee the updated value is the same as the metadata in the SGT. If not, we loop back and update \cc{PKRS} again, thus making sure the permission can only be transferred to the appointed compartment.

\textbf{\textit{P2.}} \textit{Determinism: The switch gate behavior is uniquely determined by the gate id.}\label{P2}

\begin{figure}
\begin{center}
\includegraphics[width=.9\linewidth]{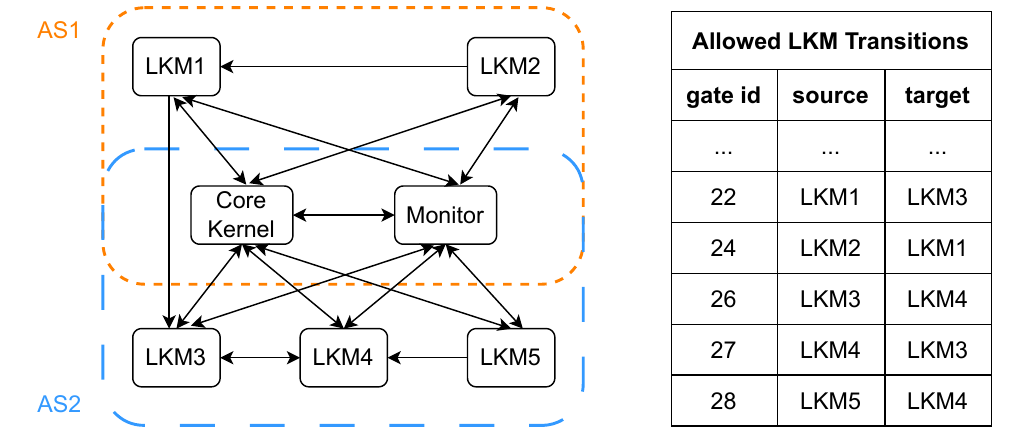}
\end{center}
\caption{\label{fig:tran} An example of compartment transition policies. The core kernel and the monitor are shared by all address spaces. AS1 contains LKM1 and LKM2, while AS2 contains LKM3-5. Edges between compartments are allowable transitions. Transitions with the core kernel and the monitor are omitted from the table.}
\vspace{-1.5em}
\end{figure}

Although the switch gate can be called from any compartment, its behavior cannot be influenced by attackers. To perform a compartment switch, the only accepted parameter is the gate id, and its behavior strictly follows the retrieved metadata. All gates must be registered first based on transition policies and do not trust the source or the target compartment. We also provide an API to register switch gates at runtime when loading new compartments. The registration requests the monitor to store metadata in the write-protected SGT. Each gate entry contains both the source and the target information as shown in \autoref{fig:switch}, including the address space (pgdir and ASID), the entry/exit address, the \cc{PKRS} value, and the private stack pointer. The entry index is used as a unique gate id. Each switch gate validates the source compartment information and atomically switches to the target compartment determined by the metadata. As a result, there is no chance for attackers to disturb the switch with compromised input, even payloads on the stack. The informative metadata guarantees the integrity of permission transfer and control flow transfer.

\autoref{fig:tran} demonstrates an example of compartment transition policies. \sys expects developers to specify flexible policies according to their requirements. Then allowed switch gates will be registered into the SGT. For instance, gate 22 allows the transition from LKM1 to LKM3, while the opposite direction (gate 23) is not allowed. If LKM5 is loaded at runtime and the user requests the monitor to register switch gates for it through our API. The monitor will check the gate to be registered against predefined policies and reject illegal requests. It guarantees that switch gates from LKM5 to modules other than LKM4 cannot be registered. As a result, an adversary cannot abuse the provided API to modify the policy maliciously.

\textbf{\textit{P3.}} \textit{Exclusivity: Compartment switches are exclusively possible via switch gates.} \label{P3}

With bi-directional isolation, the only way to access resources of other compartments is through secure compartment switches. Since switch gates are strictly based on the SGT controlled by the monitor, attackers may attempt to perform switches in other ways, such as abusing instructions for \cc{PKRS} updates. However, thanks to instruction deprivation, all instructions that can be used to break access control are validated by the monitor. As a result, switch gates exclusively guard compartment switches to support CII.

\PP{Zero-Copy Ownership Transfer}
To achieve secure and fast communication, \sys incorporates zero-copy ownership transfer for sharing data across compartments. The validation based on policies during ownership transfer thwarts confused deputy attacks, while zero-copy improves efficiency.

Data copying across compartments is a headache from multiple aspects. First, simply trusting shared data violates bi-directional isolation and leads to CIVs. Second, since the Linux kernel utilizes many tricky programming idioms like sentinel arrays and recursive structures, the synchronization between compartments requires complex and error-prone analysis and marshaling~\cite{huang2022ksplit}. Lastly, passing large buffers will cause significant performance overhead~\cite{bhattacharyya2023securecells}. Unlike PT/EPT switching-based approaches, PKS-based compartmentalization supports zero-copy communication due to its tagging mechanism.

\sys enforces single ownership of objects. As described in \textbf{\textit{I1}}~(\autoref{I1}), all kernel data is tagged with a \cc{pkey}. Besides, each compartment has its own private heap. At any time, an object is owned and can be accessed by exactly one compartment. There are three typical ways to share data: global variables, cross-compartment function call arguments and return values. Either way, access from the target compartment will trigger a page fault first. The dedicated page fault handler allows the monitor to validate the shared data to avoid confused deputy attacks. If the shared data conforms to the developer-defined policy, its \cc{pkey} will be updated to the target's, which means the ownership transfer without data copy. We provide an API to define the data transfer policy, including the source and target of shared data and its legal value ranges. So developers can deploy different compartmentalization policies flexibly and perform security checks in the page fault handler on demand.

Taking CVE-2022-1015 in \textcolor{blue}{Listing}~\autoref{l:cve} as an example, \cc{attr} is data shared between \cc{nf\_tables} and other compartments. \cc{nft\_parse\_register} parses a register value from the netlink attribute \cc{attr}. However, improper \cc{reg} validation in \cc{nft\_validate\_register\_load} could lead to integer overflow and thus out-of-bounds write issues. Based on constraints on \cc{reg} in \cc{nft\_parse\_register} and the patch for this vulnerability, we add security checks during \cc{attr} transfer. The dedicated page fault handler guarantees the register value included in \cc{attr} is within legal ranges: \cc{NFT\_REG\_VERDICT...NFT\_REG\_4} or \cc{NFT\_REG32\_00...NFT\_REG32\_15}.

\begin{listing}
\begin{center}
\begin{minted}[mathescape,
               linenos,
               numbersep=5pt,
               fontsize=\footnotesize]{C}
static int nft_validate_register_load(enum nft_registers reg, unsigned int len)
{
	if (reg < NFT_REG_1 * NFT_REG_SIZE / NFT_REG32_SIZE)
		return -EINVAL;
	if (len == 0)
		return -EINVAL;
	if (reg * NFT_REG32_SIZE + len > sizeof_field(struct nft_regs, data))
        /* A large value of reg could overflow the integer and bypass the check */
		return -ERANGE;
	return 0;
}

int nft_parse_register_load(const struct nlattr *attr, u8 *sreg, u32 len)
{
	u32 reg;
	int err;
	reg = nft_parse_register(attr);
	err = nft_validate_register_load(reg, len);
	...
}
EXPORT_SYMBOL_GPL(nft_parse_register_load);
\end{minted}
\caption{CVE-2022-1015: improper \cc{reg} validation could lead to integer overflow in net/netfilter/nf_tables_api.c.\vspace{-1em}}
\label{l:cve}
\end{center}
\vspace{-0.3em}
\end{listing}

\subsection{Two-level Compartmentalization}
\label{sec-scalability}

The Linux kernel contains thousands of LKMs developed by programmers with varying levels of expertise, requiring fine-grained compartmentalization to constrain potential vulnerabilities. However, the 4-bit \cc{pkey} limits the available compartments to a maximum of 16. Considering that we reserve \cc{pkey0} for the core kernel, \cc{pkey1} for XOM, and \cc{pkey2} for the monitor, the number of compartments available for other kernel modules is further reduced to 13, which cannot satisfy the compelling need for more isolated compartments. 

To fulfill the scalability objective~(\autoref{sec-obj2}), we utilize a two-level compartmentalization scheme. The first level is PKS-based intra-address space isolation described in \autoref{sec-bi}, while the second level is locality-aware address space switching with ASID. We group modules into multiple address spaces according to the locality of module interactions. With different memory mapping, each address space has its own 16 \cc{pkey}s, thus supporting 16 new compartments.

HAKC~\cite{mckee2022preventing} proposed a fundamentally different two-level compartmentalization approach for ARM, which does not work for \sys. Its first level is address space coloring via MTE, while the second level recycles colors through cryptographic hashes based on PA. The Intel architecture does not feature hardware primitives like PA. In addition, cryptographic operations are too expensive for frequent compartment switches. Therefore, we have to design a more compatible and efficient way to overcome the limitation of few tag bits.

\sys takes lessons from existing solutions on the userspace PKU scalability. Besides hardware extensions~\cite{schrammel2020donky, hardwarepk20}, these approaches roughly fall into two categories: dynamic PT modification~\cite{libmpk19} and multiple address spaces~\cite{gu2022epk, yuan2023vdom}. By modifying the PTE, the corresponding \cc{pkey} can be evicted and then allocated to another domain. It is essentially a kind of multiplexing and suffers from significant performance degradation with increasing domains due to TLB flushes and busy waiting. In contrast, address space switching is an efficient way to support unlimited domains. Nevertheless, PKU virtualization approaches~\cite{gu2022epk, yuan2023vdom} all rely on the underlying kernel for scheduling and management, and thus cannot migrate to the kernel space directly. MOAT~\cite{moat24} isolated BPF programs into different address spaces, but universally splitting the monolithic kernel is still a challenge.

\begin{figure}
\begin{center}
\includegraphics[width=.9\linewidth]{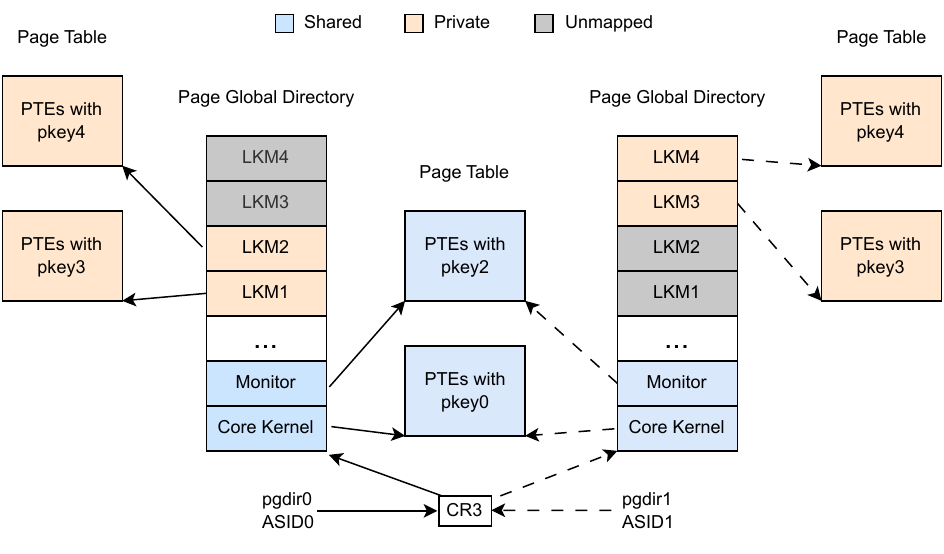}
\end{center}
\vspace{-0.8em}
\caption{\label{fig:two-level} The illustration of two-level compartmentalization. For example, LKM1 and LKM2 are isolated within the same address space through different \cc{pkey}s, while LKM1 and LKM3 are isolated in different address spaces and reuse the same \cc{pkey}.}
\vspace{-1.6em}
\end{figure}

However, we found that scalable kernel compartmentalization does not require complex address space splitting, based on two observations. First, interactions between kernel modules show a pattern of locality. Although there are thousands of LKMs in Linux, each module only needs to interact with a limited number of modules for its functionality. For example, in Linux kernel v6.1 with x86\_64 generic Kconfig, only 54 of 6296 LKMs depend on more than 12 modules, which include indirect dependencies. Thus we can split the address space based on module dependencies specified by \cc{modules.dep} to reduce address space switches. Notably, the dependencies do not mean running all dependent modules strictly at the same time. With more address space switching, we can support unlimited compartments as a kind of multiplexing, even for the 54 cases. In \autoref{fig:two-level}, we assume that LKM2 depends on LKM1 and map them in the same address space, while the unrelated modules like LKM3 and LKM4 can be isolated in another address space to reuse \cc{pkey}s.

Second, the Linux kernel employs hierarchical PTs for memory mapping and we do not have to duplicate all the PTs for multiple address spaces. \autoref{fig:two-level} shows an example with two-level PTs. We divide the PTs into the shared and private parts. Since all modules need to interact with the core kernel and the monitor, all address spaces share the same mapping of them. On the other hand, the memory reusing the same \cc{pkey}s with other address spaces is private to prevent collision. During address space switching, we unmap the source space's private part and remap the private part of the target. Thanks to the hierarchical structure, PT updates mainly occur at the low-level PTs. Thus we can adjust the memory mapping efficiently to maintain different views. Besides, we attach each address space with an address space identifier (ASID), also called process context identifier (PCID) for x86, to avoid TLB flushes as optimization. The address translation only uses TLB entries with the same ASID as the current page table base register (\cc{CR3}). Specifically, address space switches are also advised by the monitor based on the SGT. The registered gate entry records the source and target page global directory addresses (\cc{pgdir}) as well as ASIDs (\autoref{fig:switch}). Due to the secure properties of gates, attackers cannot forge a malicious address space to bypass isolation. Thus, we realize secure and locality-aware two-level compartmentalization that supports unlimited compartments.

\section{Implementation}
\label{sec5}

To test the effectiveness of our design, we have implemented a prototype based on the Linux kernel v6.1, the newest long-term support (LTS) version when we started this work, and LLVM version 14.0.0. The \sys system consists of a set of kernel patches, a lightweight monitor module, and LLVM passes for instruction deprivation and automated gate instrumentation. Since the Linux kernel did not have support for Intel PKS at the time of writing, we implemented a series of patches to provide PKS-related API like existing API for PKU, such as \cc{pkey} allocation and \cc{PKRS} updates. This part includes 869 lines of addition and 90 lines of deletion. On the basis of these APIs, we implemented the in-kernel monitor as a separate LKM for portability, which only contains 2759 lines of code. The small codebase makes formal modeling and verification possible~\cite{formal22}, which is one of our ongoing works.

Although \sys is implemented over Linux for its open-source nature, we expect the main design principles like bi-directional isolation and two-level compartmentalization could also be extended to other commodity OSs with PKS support. Here we highlight some key points.

\PP{PKRS State}
\sys manages the \cc{PKRS} state by the monitor. Unlike \cc{PKRU}, the per-thread \cc{PKRS} register is not XSAVE-supported by hardware. Therefore, our software implementation needs to save and restore its state on context switches and exceptions. However, attackers could overwrite the stored \cc{PKRS} in memory to gain control over the register indicating access permissions. As a countermeasure, we store the \cc{PKRS} state in a specific region write-protected by the monitor, which ensures that attackers cannot tamper with its value.

\PP{Multi-threading Support}
\sys supports multi-threading securely. Since \cc{PKRS} is per-thread and its state is write-protected, the access control inherently supports multi-threading. Concurrent access to shared data may lead to vulnerabilities, such as time-of-check to time-of-use (TOCTTOU) attacks. In response, we enforce exclusive access to shared data through single ownership, which means that concurrent compartments other than the owner cannot modify the data.

\PP{Write-Protected Page Tables}
We reserve a page pool tagged with the monitor's \cc{pkey} for PTs. To solve the page split issue of the direct map region, the pool remains enough pages to break down huge pages into the 4 KB size. Then we identify all kernel functions that allocate or update PTs. For PT allocation, a wrapper function is added to allocate from the reserved page pool. For PT updates, we insert switch gates to the monitor to access the PT securely. In order to prevent omissions, we apply a checker that scans for unprotected PT pages. PTs that are allocated before \sys initialization are also caught by the checker and then tagged with the \cc{pkey}. As a result, we guarantee all PTs are write-protected by the monitor.

\PP{Private Heap}
We implement private heaps to protect dynamically allocated objects of compartments. Similarly to the PT, dedicated memory caches with specific \cc{pkey} are provided for private object allocation. We extend both the buddy allocator and the SLAB/SLUB allocator to return objects tagged with the desired \cc{pkey} from the caches.

\PP{Page Alignment}
Since \cc{pkey}s are associated with PTEs, \sys can only protect memory at the page granularity, which is generally 4KB. Objects may interleave with other compartments across the page boundaries and lead to imprecise access control. To prevent this, we force all compartments to allocate objects from their private heaps. The reserved page pool ensures the compartment memory is page-aligned. 

Data sharing across compartments also poses a challenge. As described in \autoref{I3}, zero-copy data transfer performs at the granularity of a page, which means that potentially more data gets shared and thus increases attack surfaces. One solution is to include only the objects to be shared on the transferred page. However, it may waste a lot of memory. As a trade-off between security and memory overhead, we group shared objects into different privilege classes according to the source and the target compartments, then put objects with the same privilege on a page. For instance, all objects shared between LKM1 and LKM2 could be gathered on the same page, while objects shared between LKM1 and LKM3 are located on another page. With this kind of object grouping, even over-sharing does not give a compartment access to resources it should not have.

\PP{Switch Gate Registration and Instrumentation}
Switch gates are registered in pairs for bi-directional isolation. If a gate id $x$ represents the entry to a specific compartment, then $x + 1$ represents the return, and corresponding SGT entries contain the exact opposite of source and target information. Thus, the paired gates enforce CII and thwart ROP-like attacks on cross-compartment calls. \sys requires the trusted policy developers to define a set of legal transitions like \autoref{fig:tran}. During the initialization, we register two special gates as the entry and exit of the monitor. After that, the SGT is exclusively owned by the monitor and all registrations should transfer to it first. Then, the monitor will check the gate metadata against the policy and add the validated one to the SGT. Although users may attempt to register gates at runtime via an API for newly loaded compartments, these requests are also guarded by the monitor from violating the prior policy. To perform registered switches, the insertion of switch gate calls with the gate id is needed. For usability, we implement an LLVM pass that automatically inserts switch gate registrations and switch gate calls at a list of interfaces based on the boundary analysis.

\subsection{Use Case}
\label{sec5-1}
We choose LKM isolation as the use case of \sys for security and practicability reasons. First, LKMs cause the majority of vulnerabilities as shown in \autoref{fig:vul}. Second, diverse LKMs provide specialized functionality, which is suitable for privilege separation. Third, the development conventions make LKM boundaries relatively clear, and we can benefit from existing static analysis methods to perform automated isolation via LLVM passes.

For compartmentalization policy generation, we apply several analyses on the LLVM intermediate representation (IR) of the kernel. The LKM's interfaces and shared data are identified by KSplit~\cite{huang2022ksplit} static analysis, while security checks are specified by a constraint analysis~\cite{lu19missing-check}. Specifically, a boundary analysis will collect all interface functions between the LKM and other kernel compartments, including exported symbols, registered function pointers, and interrupt handlers. It is a bi-directional analysis and the interfaces will be classified as entries or exits. Then we identify the data accessible from both sides of the isolation boundaries. This shared data analysis is performed on the program dependence graph (PDG) built by SVF~\cite{svf16}. For data validation, we collect constraints on shared data along the use-define chains from interfaces and generate security checks for them. With these boundaries, shared data, and security checks, we can insert switch gates for secure switching and cross-compartment communication.

Given a specific LKM that needs to be isolated, we first allocate a \cc{pkey} and tag its memory with the \cc{pkey} during the module initialization. Then we register and insert the LKM entry/exit gates according to policies before loading it into the kernel. Different from KSplit, \sys employs zero-copy ownership transfer and thus does not need complex synchronization for data sharing. As a result, we write a simple LLVM pass instead of a specific interface definition language (IDL) compiler to realize automated LKM isolation with PKS.

\section{Evaluation}
\label{sec6}
This section answers the following questions through comprehensive evaluations of \sys for security, performance, and scalability:

\noindent \textbf{RQ1:} Can \sys defend against potential attacks under the proposed threat model (\autoref{sec6-1})?

\noindent \textbf{RQ2:} How much overhead does \sys introduce to micro-benchmarks (\autoref{sec6-2-1})?

\noindent \textbf{RQ3:} What performance overhead does \sys impose on real-world applications (\autoref{sec6-2-2})?

\noindent \textbf{RQ4:} Is the performance of \sys scalable for multiple compartments (\autoref{sec6-2-3})?

\noindent \textbf{RQ5:} What's the memory overhead of \sys (\autoref{sec-mem})?

\subsection{Security Analysis}
\label{sec6-1}
We assess how \sys enforces security invariants \textbf{\textit{I1}}-\textbf{\textit{I3}} (\autoref{sec-invariant}) against attack vectors under our threat model for \textbf{\textit{RQ1}}. Furthermore, to evaluate the effectiveness of \sys in real-world situations, we performed some penetration tests and investigated real vulnerabilities as case studies.

\subsubsection{Data-oriented Attacks}
\label{sec6-1-1}
Attackers exploiting vulnerabilities within a compartment may attempt to get arbitrary read-and-write primitive and corrupt data of other compartments. \sys thwarts this threat by tagging compartments' memory with different \cc{pkey}s and thus enforces data integrity~(\textbf{\textit{I1}}). The \cc{PKRS} indicates the current compartment's permission and only allows access to its own data, adhering to the principle of least privilege. Since the access control by \cc{PKRS} is thread-local, \sys also prevents cross-thread attacks inherently. We comprehensively isolate all types of kernel memory regions, including the heap, stack, physmap, and MMIO regions. The private heap guarantees the single ownership of heap objects, while the private stack hides the execution context from other compartments. PKS hardware validation also works on the physmap and MMIO regions, while malicious direct memory access (DMA) actions~\cite{dma21} are guarded by IOMMU~\cite{dautenhahn2015nested}. With bi-directional isolation, even vulnerabilities in the core kernel cannot be exploited to compromise other subsystems.

Attackers may also seek ways to bypass or disable the PKS-based protection, either tampering with the permission bits in PTs, or abusing control registers. We block these attempts by write-protected PTs and instruction deprivation. PT updates and sensitive instructions can only be performed by the monitor with validation. Besides, the \cc{PKRS} state is also exclusively managed by the monitor. Notably, PKU pitfalls~\cite{pitfall20, voulimeneas2022you, schrammel2022jenny} like syscall abuses are under a weaker threat model focusing on the userspace. The low-level countermeasures we take for the privileged kernel will make these attacks no sense.

\subsubsection{Control Flow Hijacking}
\label{sec6-1-2}
In addition to data-oriented attacks, control flow hijacking is another traditional threat to kernel security. Besides the private stack mentioned above, we mitigate it broadly in a lightweight but effective way, XOM (\textbf{\textit{I2}}). On the basis of the $W\bigoplus X$ policy, all kernel code regions are unwritable and unreadable. Therefore, attackers cannot inject malicious code directly, and reusing the existing code also becomes difficult without layout information. With existing diversification schemes like KASLR and data compartmentalization, it is hard to break XOM through memory disclosure. Instead of expensive whole kernel CFI, we focus on the cross-compartment calls and enforce CII (\textbf{\textit{I3}}). The in-kernel monitor manages the SGT to make switches atomic, deterministic, and exclusive, which prevents attackers from abusing the gates to jump to illegal compartments and gain elevated permissions.

\subsubsection{Confused Deputy Attacks}
\label{sec6-1-3}
Compartmentalization faces a special challenge of confused deputy attacks at the compartment interfaces. To address this problem, we highlight bi-directional isolation under the mutually untrusted threat model and enforce CII (\textbf{\textit{I3}}) with the help of developer-specified policies. Specifically, with distrust in mind, the monitor will check both the source and the target metadata at interfaces. Cross-compartment communication is performed in the way of zero-copy ownership transfer. Any compartmentalization policy violation will be detected in the page fault handler.

We trust the system developer who specifies prior policies. Although an adversary may attempt to register malicious switch gates at runtime for newly loaded compartments, these requests are also guarded by the monitor from violating the policy.

\subsubsection{Penetration Tests}
We instantiated the above attack vectors on the LKM isolation use case to test the security of \sys. We simulate an attacker to (1) modify the heap object of other compartments, (2) tamper the PT directly, (3) forge PTs by \cc{mov-to-CR3}, (4) update \cc{PKRS} directly, (5) abuse the switch gate to hijack control flow, and (6) pass malicious data through interfaces. All attempts are detected by protection key violation or rejected by the monitor, which shows that \sys is immune to these attacks.

\begin{table}
\footnotesize
\begin{center}
\resizebox{\columnwidth}{!}{%
\begin{tabular}{l|l|l|l}
\hline
\textbf{CVE ID} & \textbf{Root Cause}                                                                                    & \textbf{Compartment} & \textbf{Countermeasures}                                                                                                                       \\ \hline
2023-4147  & \begin{tabular}[c]{@{}l@{}}use-after-free in\\ net/netfilter/nf\_tables\_api.c\end{tabular}            & nf\_tables           & \multirow{6}{*}{\begin{tabular}[c]{@{}l@{}}~\\~\\~\\~\\The private heap prevents\\ the compartment from\\ corrupting other kernel\\ objects.\end{tabular}} \\ \cline{1-3}
2022-24122 & \begin{tabular}[c]{@{}l@{}}use-after-free in\\ kernel/ucount.c\end{tabular}                            & core kernel          &                                                                                                                                                \\ \cline{1-3}
2022-27666 & \begin{tabular}[c]{@{}l@{}}heap out-of-bounds write in\\ net/ipv6/esp6.c\end{tabular}                  & esp6                 &                                                                                                                                                \\ \cline{1-3}
2022-25636 & \begin{tabular}[c]{@{}l@{}}heap out-of-bounds write in\\ net/netfilter/nf\_dup\_netdev.c\end{tabular}  & nf\_dup\_netdev      &                                                                                                                                                \\ \cline{1-3}
2021-22555 & \begin{tabular}[c]{@{}l@{}}heap out-of-bounds write in\\ net/netfilter/x\_tables.c\end{tabular}        & x\_tables            &                                                                                                                                                \\ \cline{1-3}
2018-5703  & \begin{tabular}[c]{@{}l@{}}heap out-of-bounds write in\\ net/ipv6/tcp\_ipv6.c\end{tabular}             & ipv6                 &                                                                                                                                                \\ \hline
2023-0179  & \begin{tabular}[c]{@{}l@{}}stack buffer overflow in\\ net/netfilter/nft\_payload.c\end{tabular}        & nf\_tables           & \multirow{2}{*}{\begin{tabular}[c]{@{}l@{}}The private stack blocks\\ cross-compartment stack\\ corruption.\end{tabular}}                      \\ \cline{1-3}
2018-13053 & \begin{tabular}[c]{@{}l@{}}integer overflow in\\ kernel/time/alarmtimer.c\end{tabular}                 & core kernel          &                                                                                                                                                \\ \hline
2022-1015  & \begin{tabular}[c]{@{}l@{}}improper input validation in\\ net/netfilter/nf\_tables\_api.c\end{tabular} & nf\_tables           & \multirow{3}{*}{\begin{tabular}[c]{@{}l@{}}~\\The monitor-enforced\\ interface checks thwart\\ confused deputy attacks.\end{tabular}}             \\ \cline{1-3}
2022-0492  & \begin{tabular}[c]{@{}l@{}}missing authorization in\\ kernel/cgroup/cgroup-v1.c\end{tabular}           & core kernel          &                                                                                                                                                \\ \cline{1-3}
2017-18509 & \begin{tabular}[c]{@{}l@{}}improper input validation\\ in net/ipv6/ip6mr.c\end{tabular}                & ipv6                 &                                                                                                                                                \\ \hline
\end{tabular}%
}
\end{center}
\caption{Representative Linux kernel CVEs, their root causes, the located compartment, and the countermeasures of \sys.}
\label{tb:cve}
\vspace{-2em}
\end{table}

\subsubsection{Real-world Vulnerabilities}
To evaluate how \sys mitigates real-world vulnerabilities, we select 11 representative Linux kernel CVEs according to the following criteria: (1)~they should contain multiple error types and cover attack vectors we considered in \autoref{sec6-1-1} - \autoref{sec6-1-3}; (2)~they should be distributed in diverse compartments, including the core kernel and other LKMs; (3)~their Proof-of-Concept (PoC) programs or exploits are available; (4)~we prefer recent vulnerabilities with high severity, i.e., CVSS 3.x rating \textit{7.8} or higher except for CVE-2018-13053 (\textit{3.3}) and CVE-2022-1015 (\textit{6.6}). \autoref{tb:cve} illustrates the CVE set with their root causes, the located compartment, and the countermeasures.

Since the compartment can only access its own objects tagged with the corresponding \cc{pkey}, the private heap confines the damage of both heap use-after-free (UAF) vulnerabilities like CVE-2023-4147 and heap out-of-bounds (OOB) vulnerabilities like CVE-2022-27666. Stack buffer overflow (CVE-2023-0179) or integer overflow (CVE-2018-13053) can be exploited to corrupt stack variables or tamper with the return addresses. \sys mitigates these threats by maintaining a private stack for each compartment. In general, the data integrity (\textbf{\textit{I1}}) blocks any attempts of memory corruption, which is the dominant threat to the Linux kernel. Vulnerabilities with insufficient parameter validation lead to confused deputy attacks, which is a challenge for traditional data protection techniques~\cite{lu2023practical}. We address this problem through bi-directional isolation. The in-kernel monitor checks both the source and the target information during switching to enforce CII~(\textbf{\textit{I3}}). Thus, only legal parameters can be transferred across compartments. For example, \sys will strictly validate the input register values of \cc{nft\_parse\_register\_load} function to mitigate CVE-2022-1015 (\textcolor{blue}{Listing} \autoref{l:cve}). Besides, many listed vulnerabilities, such as CVE-2021-22555, CVE-2022-1015, and CVE-2022-25636, may be further exploited to launch ROP attacks. As a principled countermeasure, XOM (\textbf{\textit{I2}}) makes it extremely hard to find usable gadgets. Overall, \sys provides comprehensive protection against various kinds of real-world vulnerabilities to fulfill security objectives (\autoref{sec-obj1}).

\subsection{Performance Evaluation}
\label{sec6-2}
We first describe our experiment setup. All evaluations were conducted on a machine with Intel Core i7-12700H CPU, 16 GB memory, and 500GB disk, running Ubuntu-22.04 with Linux kernel v6.1. Further, we set the kernel to performance mode and locked the CPU frequency to avoid randomness.

There are five types of configuration: (1) \textit{monitor}, which only enables \sys in-kernel monitor for the PT and SGT protection; (2) \textit{ipv6} with the compartmentalized \cc{ipv6} module; (3) \textit{ipv6-nft}, which isolates \cc{ipv6} and \cc{nf\_tables} into separate compartments; (4) \textit{lkm-20}, which isolates 20 LKMs into different address spaces detailed in \autoref{sec6-2-3}; (5) \textit{lkm-160}, which compartmentalizes all 160 LKMs with \cc{localmodconfig} on the experimental machine. We specify compartmentalization policies for each setting based on the boundary analysis, shared data analysis, and security check analysis described in \autoref{sec5-1}.

\subsubsection{Micro-benchmarks}
\label{sec6-2-1}
For \textbf{\textit{RQ2}}, we measure the CPU cycles of the compartment switch and then use LMbench~\cite{mcvoy1996lmbench} to evaluate the latency and bandwidth overhead of \sys on micro-operations.

\begin{table}[htbp]
\footnotesize
\begin{center}
\begin{tabular}{@{}lr@{}}
\toprule
\textbf{Operation}                   & \textbf{Cost (cycles)} \\ \midrule
\cc{PKRS} update by \cc{wrmsr}                 & 185.31                 \\
compartment switch                   & 224.31                 \\
com-switch without stack switch      & 203.96                 \\
com-switch with address space switch & 523.69                 \\
syscall null                         & 293.86                 \\
hypercall null                       & 2894.06                \\
\cc{vmfunc} (EPT switch)                  & 198.82                 \\ \bottomrule
\end{tabular}
\end{center}
\caption{Cost comparison of \sys micro-operations.}
\label{tb:cycles}
\vspace{-1em}
\end{table}

We list the cost of the compartment switch operations in \autoref{tb:cycles}. The CPU cycles are measured by the \cc{rdtscp} instruction, and we average ten runs of 10000 invocations. A single \cc{PKRS} update by \cc{wrmsr} takes about 185.31 cycles, while a secure compartment switch through the gate takes 224.31 cycles. As a reference, the null syscall on the experimental machine costs 293.86 cycles, and a simple hypercall into the hypervisor costs 2894.06 cycles. Besides, the \cc{PKRS} update is even faster than the \cc{vmfunc} instruction, which is well-known for its efficient EPT switch. Considering the security guarantees, virtualization-based approaches will suffer worse performance overhead due to nested paging and I/O virtualization. Although the additional address space switch increases the cost to 523.69 cycles, the locality-aware two-level compartmentalization scheme makes this case rarely happen and almost does not affect performance, which is shown in the following evaluation.

\begin{table}[htbp]
\footnotesize
\begin{center}

\resizebox{\columnwidth}{!}{%
\begin{tabular}{@{}lrrrrr@{}}
\toprule
\textbf{Benchmarks}  & \textbf{monitor} & \textbf{ipv6} & \textbf{ipv6-nft} & \textbf{lkm-20}   & \textbf{lkm-160} \\ \midrule
\multicolumn{6}{c}{\cellcolor[HTML]{EFEFEF}\textit{\textbf{Latency}}}                                                                                                                                                                   \\
syscall null                            & -0.37                                & -0.28                             & -0.30                                 & 0.09 \textit{(4)}                   & 0.36 \textit{(12)}                            \\
simple read                             & -2.65                                & -1.68                             & -1.42                                 & -1.92 \textit{(4)}                           & -1.90 \textit{(12)}                           \\
simple write                            & -2.30                                & -2.34                             & -2.08                                 & -2.00 \textit{(4)}                           & -1.58 \textit{(12)}                           \\
simple stat                             & -0.03                                & 0.58                              & 0.05                                  & -0.17 \textit{(4)}                           & 0.80 \textit{(12)}                            \\
simple fstat                            & 0.22                                 & -0.74                             & -1.20                                 & -0.74 \textit{(4)}                           & -0.51 \textit{(12)}                           \\
open/close                              & 0.85                                 & 0.87                              & 0.79                                  & 0.07 \textit{(4)}                            & 1.24 \textit{(12)}                            \\
select on fd's                          & -1.26                                & -1.20                             & -1.38                                 & -0.83 \textit{(4)}                           & -1.18 \textit{(12)}                           \\
select on tcp fd's                      & -0.64                                & -0.41                             & -0.57                                 & -0.30 \textit{(4)}                           & -0.53 \textit{(12)}                           \\
signal install                          & -0.01                                & -0.19                             & 0.39                                  & 0.50 \textit{(4)}                            & 0.43 \textit{(9)}                             \\
signal handler                          & -0.21                                & -0.26                             & 1.53                                  & -1.96 \textit{(4)}                           & -0.14 \textit{(9)}                            \\
proc fork+exit                          & 26.88                                & 27.10                             & 27.79                                 & 27.84 \textit{(4)}                           & 28.30 \textit{(9)}                            \\
proc fork+exec                          & 15.86                                & 15.80                             & 16.06                                 & 15.86 \textit{(4)}                           & 16.26 \textit{(9)}                            \\
proc shell                              & 14.59                                & 14.47                             & 14.77                                 & 14.63 \textit{(4)}                           & 15.25 \textit{(9)}                            \\
page fault                              & 39.71                                & 39.67                             & 39.76                                 & 39.89 \textit{(6)}                           & 38.64 \textit{(9)}                            \\
pipe latency                            & 0.68                                 & 1.09                              & 1.77                                  & 1.02 \textit{(4)}                            & 1.89 \textit{(11)}                            \\
UDP latency                             & 0.96                                 & 0.78                              & 1.41                                  & 2.50 \textit{(4)}                            & 3.09 \textit{(11)}                            \\
TCP latency                             & 0.35                                 & 0.62                              & 1.31                                  & 1.57 \textit{(6)}                            & 3.54 \textit{(11)}                            \\
\multicolumn{6}{c}{\cellcolor[HTML]{EFEFEF}\textit{\textbf{Bandwidth}}}                                                                                                                                                                 \\
file write bandwidth                    & -1.75                                & -3.99                             & -3.57                                 & -2.78 \textit{(6)}                           & -3.05 \textit{(9)}                            \\
pipe bandwidth                          & 1.42                                 & -0.35                             & 0.77                                  & -0.11 \textit{(4)}                           & 0.18 \textit{(11)}                            \\
AF\_UNIX bandwidth                      & 0.34                                 & 0.58                              & 0.23                                  & 0.35 \textit{(4)}                            & 0.23 \textit{(11)}                            \\ \bottomrule
\end{tabular}%
}
\end{center}
\caption{\sys performance overhead (in \% over the vanilla kernel) on LMbench. The numbers in parentheses represent the number of compartments traversed for each benchmark.}
\label{tb:lmbench}
\vspace{-2em}
\end{table}

\autoref{tb:lmbench} illustrates the LMbench evaluation results, with vanilla kernel v6.1 as the baseline. To achieve stable results, we run each benchmark 100 times and take the average. 

The table reveals two main observations. Focusing on the columns, \sys introduces negligible overhead of less than 1\% in most benchmarks, except for the process operations and page faults. The increased runtime overhead stems from compartment switches with the monitor due to the PT protection. However, as we will see from the macro-benchmarks, it does not have a noticeable impact on real-world applications. While looking at the table by rows, we can observe that the system-wide micro-benchmark is almost unaffected by LKM isolation, regardless of the number of involved compartments. This is because the switches between LKMs occur rarely on LMbench. Especially, the results of \textit{lkm-20} and \textit{lkm-160} show the scalability of \sys (\textbf{\textit{RQ4}}). A few benchmarks perform slightly better than the vanilla kernel within a reasonable margin of fluctuation. It comes down to the system noise like improved cache hit rates, which is hard to avoid as prior works~\cite{iskios21, dope23, sok19benchmark} show.

\subsubsection{Macro-benchmarks}
\label{sec6-2-2}
We performed two sets of evaluations to answer \textbf{\textit{RQ3}}. First, we demonstrate the system-wide impact of \sys by running a collection of Phoronix Test Suites~\cite{phoronix}. Then, we zoom in on the performance overhead of a specific compartment. We select the vulnerable \cc{ipv6} module as the target and test it on ApacheBench~\cite{ab} so that we can draw a comparison with HAKC - one of the state-of-the-art kernel compartmentalization approaches.

The Phoronix suites provide a large number of system-wide tests, from which we select some representative ones to comprehensively characterize the performance of \sys. Our benchmarks are split into application tests and stress tests for specific subsystems. The application benchmarks include nginx (measures sustained requests/second, varying the number of concurrent connections); phpbench (tests the PHP interpreter); pybench (tests basic, low-level functions of Python); povray (3D ray tracing); gnupg (encryption time with GnuPG). Stress tests include dbench (measures disk performance via file system calls, varying the number of clients); postmark (transactions on 500 small files simultaneously); sysbench (performs CPU and memory tests). \autoref{tb:phoronix} shows the performance overhead compared to the vanilla kernel v6.1. Nginx incurs 3.57\%-7.29\% slowdown since the continuous network packet processing results in frequent compartment switches. All other benchmarks present negligible overhead and the average performance overhead with the \cc{ipv6} module compartmentalized is 1.28\%.

\begin{table}[t]
\footnotesize
\begin{center}
\resizebox{\columnwidth}{!}{%
\begin{tabular}{@{}lrrrrr@{}}
\toprule
\textbf{Benchmarks} & \textbf{monitor} & \textbf{ipv6} & \textbf{ipv6-nft} & \textbf{lkm-20}   & \textbf{lkm-160}      \\ \midrule
nginx-100           & 4.88             & 5.03          & 6.01              & 5.70 \textit{(7)}          & 7.29 \textit{(19)}             \\
nginx-200           & 4.47             & 4.55          & 5.54              & 5.38 \textit{(7)}          & 6.54 \textit{(19)}             \\
nginx-500           & 3.57             & 3.68          & 4.40              & 4.51 \textit{(7)}          & 5.74 \textit{(19)}             \\
phpbench            & -0.24            & -0.12         & -0.44             & -0.28 \textit{(7)}         & 0.33 \textit{(18)}             \\
pybench             & 0.35             & 0.17          & 0.43              & 0.52 \textit{(7)}          & 1.37 \textit{(18)}             \\
povray              & 0.16             & 0.57          & 0.22              & 0.39 \textit{(7)}          & 0.2 \textit{(17)}              \\
gnupg               & 0.10             & 0.01          & 0.35              & 0.08 \textit{(7)}          & 1.03 \textit{(18)}             \\
dbench-1            & 0.19             & 0.20          & 0.19              & 0.04 \textit{(7)}          & 0.47 \textit{(19)}             \\
dbench-48           & 0.52             & 1.05          & 1.73              & 3.74 \textit{(7)}          & 5.61 \textit{(19)}             \\
dbench-256          & 0.22             & 1.22          & 2.38              & 1.64 \textit{(7)}          & 2.11 \textit{(19)}             \\
postmark            & 1.84             & 0.00          & 1.14              & 1.14 \textit{(7)}          & 0.39 \textit{(18)}             \\
sysbench-cpu        & -0.05            & -0.03         & -0.04             & -0.01 \textit{(7)}         & 0.01 \textit{(19)}             \\
sysbench-mem        & 0.02             & 0.26          & -0.53             & 0.53 \textit{(7)}          & 0.69 \textit{(18)}             \\
\textbf{Average}    & \textbf{1.23}    & \textbf{1.28} & \textbf{1.64}     & \textbf{1.80 \textit{(7)}} & \textbf{2.44 \textit{(18.46)}} \\ \bottomrule
\end{tabular}%
}
\end{center}
\caption{\sys performance overhead (in \% over the vanilla kernel) on Phoronix Test Suites. The numbers in parentheses represent the number of compartments traversed for each benchmark.}
\label{tb:phoronix}
\vspace{-1em}
\end{table}

\begin{table}[t]
\footnotesize
\begin{center}
\resizebox{\columnwidth}{!}{%
\begin{tabular}{@{}lll|l|l@{}}
\toprule
\multicolumn{3}{c|}{\textbf{net}} & \multicolumn{1}{c|}{\textbf{fs}} & \multicolumn{1}{c}{\textbf{usb}} \\ \midrule
nfnetlink        & nf\_conntrack & ip\_tables  & overlay      & typec              \\
nf\_defrag\_ipv4 & nf\_nat       & ip6\_tables & pstore\_zone & tps6598x           \\
nf\_defrag\_ipv6 & nf\_tables    & nft\_compat & ramoops      & xhci\_pci\_renesas \\
ipv6             & x\_tables     & nft\_nat    & pstore\_blk  & xhci\_pci          \\ \bottomrule
\end{tabular}%
}
\end{center}
\caption{The 20 compartmentalized LKMs from the 3 most vulnerable subsystems.}
\label{tb:lkm}
\vspace{-2.5em}
\end{table}

\begin{figure}[t]
\begin{center}
\includegraphics[width=.85\linewidth]{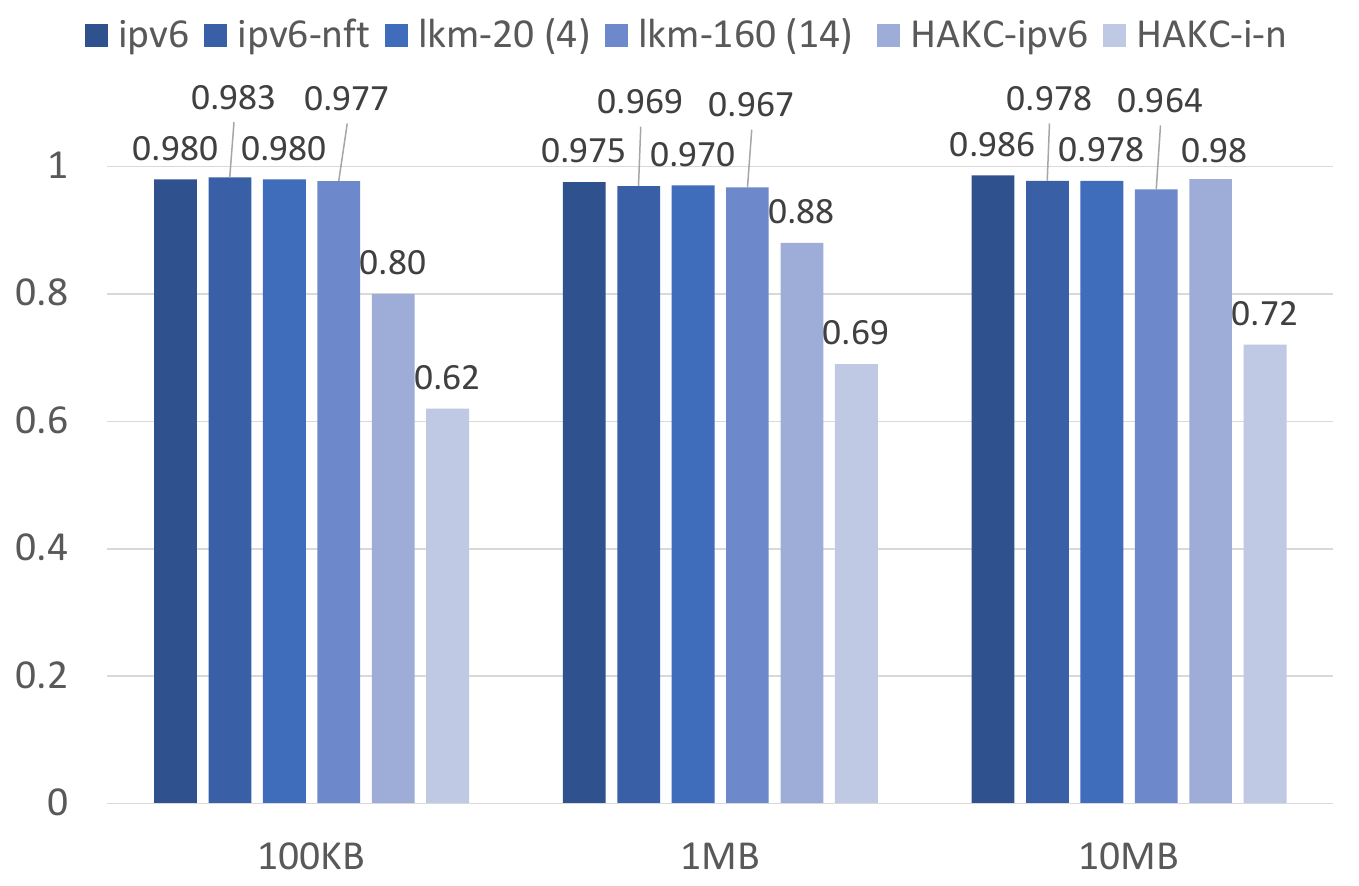}
\end{center}
\vspace{-0.5em}
\caption{\label{fig:abr} \sys performance overhead normalized to the vanilla kernel when transferring various sized payloads on ApacheBench (requests/sec), compared with the overhead of HAKC~\cite{mckee2022preventing}.}
\vspace{-1em}
\end{figure}

To further measure the overhead on the isolated \cc{ipv6} module. We use ApacheBench to retrieve a 100KB, 1MB, and 10MB file 1000 times from a local Apache server through an ipv6 address. The \cc{nf\_tables} module implements a packet filtering mechanism within the kernel. Each experiment is repeated 10 times to avoid randomness.  The results normalized to the vanilla kernel are listed in \autoref{fig:abr}, and the metric of transfer rate reveals a similar trend to requests/second. We also show the overhead reported in HAKC for comparison. The overhead of most settings is around 2\%, which is much better than HAKC. Even with more compartments traversed, the performance does not show a significant degradation. Specifically, the \textit{lkm-20} setting executes 4 compartments, while the \textit{lkm-160} setting executes 14 compartments. In contrast, HAKC only evaluated the \cc{ipv6} and \cc{nf\_tables} modules.

\subsubsection{Scalability}
\label{sec6-2-3}

One of the major drawbacks of the previous work is the significant degradation of performance as the number of isolated domains increases. In particular, HAKC suffers from a linear growth of 14\%-19\% per compartment involved in overhead, which makes it impractical. To indicate the effectiveness of locality-aware two-level compartmentalization and answer \textbf{\textit{RQ4}}, we perform two sets of experiments. 

\textit{lkm-20} selects LKMs listed in \autoref{tb:lkm} from the 3 most vulnerable subsystems according to the vulnerability analysis in \autoref{sec2-vul}. There are 12 LKMs from \cc{net}, 4 LKMs from \cc{fs}, and 4 LKMs from \cc{usb}. The \cc{net} LKMs belong to one address space, while other LKMs belong to the other. \textit{lkm-160} extensively isolate all LKMs supported by the experimental machine with \cc{localmodconfig}, 160 compartments in total. We partition these LKMs into different address spaces according to module dependencies, specify compartmentalization policies, and comprehensively isolate them with \sys. If a module depends on more than 12 modules, the dependent modules will occupy more than one address space and \sys will perform address space switching on demand. The evaluation results in \autoref{tb:lmbench}, \autoref{tb:phoronix}, and \autoref{fig:abr} show that the increase in the number of compartments and address spaces does not significantly increase overhead, benefiting from our lightweight switch gates and locality-aware design. 

Since not all compartments present are executed, we recorded the number of compartments actually traversed for each benchmark to reflect scalability faithfully. The results reveal several findings. First, each benchmark involves only a few compartments, which also corroborates our locality-aware design. Take Phoronix test suites for example, the \textit{lkm-20} setting executes 7 compartments, and the \textit{lkm-160} setting executes 18.46 compartments on average. Nevertheless, we have to prepare a large number of compartments for complex workloads so as to confine all possible vulnerabilities. Second, the presence of compartments not involved does not affect performance. Third, looking at \autoref{tb:lmbench} and \autoref{tb:phoronix} by rows, the increase in the number of traversed compartments only slightly adds to the overhead. Lastly, the vast majority of compartment transitions are monitor entries/exits for privileged instruction or critical objects (e.g., PT) updates, which explains the slight overhead caused by other traversed compartments. For LMbench, 99.99\% of \textit{lkm-20} transitions and 99.75\% of \textit{lkm-160} transitions are monitor entries/exits, while for Phoronix, the percentages are 99.82\% of \textit{lkm-20} transitions and 93.40\% of \textit{lkm-160} transitions. Overall, with two-level compartmentalization, we can support unlimited compartments without sacrificing performance.

\subsection{Memory Overhead}
\label{sec-mem}

\begin{figure}[t]
\begin{center}
\includegraphics[width=.93\linewidth]{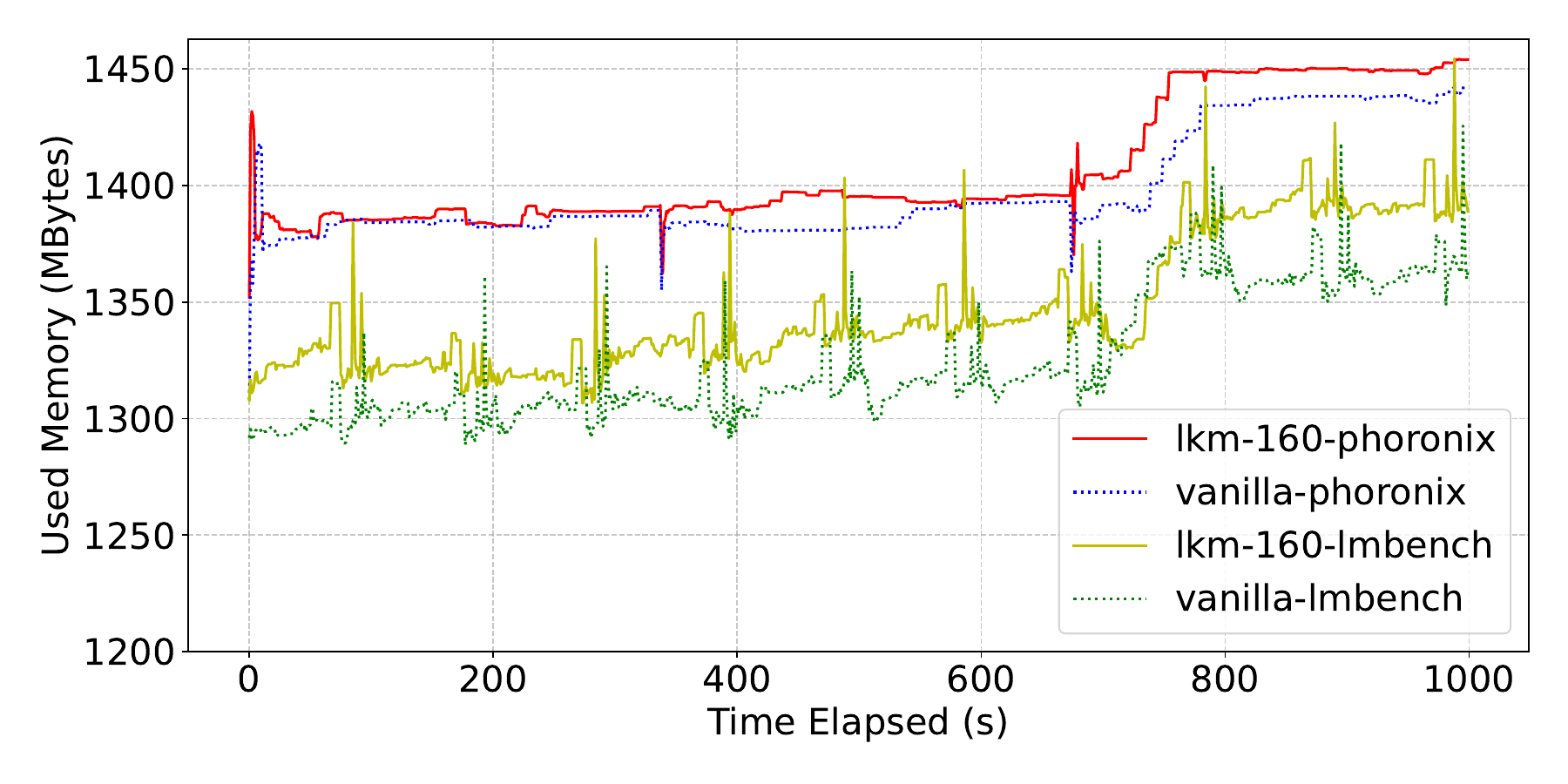}
\end{center}
\vspace{-1em}
\caption{\label{fig:mem} Memory usage of \sys when running LMbench and Phoronix with \textit{lkm-160} and the vanilla kernel.}
\vspace{-2em}
\end{figure}

As mentioned in \autoref{sec5}, \sys groups shared objects based on privilege, and makes them page-aligned to avoid over-sharing between compartments. For \textbf{\textit{RQ5}}, we measure the memory overhead caused by page alignment. \autoref{fig:mem} presents the memory usage when running LMbench and Phoronix with \textit{lkm-160} and the vanilla kernel. On average, the memory overhead is 1.66\% for LMbench and 0.63\% for Phoronix, which is negligible for modern systems.

\section{Discussion}
\label{sec7}

\PP{PKS Granularity}
\sys leverages PKS for compartmentalization, which only supports page-size granularity. To prevent privilege escalation caused by over-sharing and imprecise access control, we group shared objects into different privilege classes according to the source-target compartments and make all compartment memory page-aligned. We argue that the page granularity is already considered fine-grained for monolithic kernel compartmentalization and is comparable to other PT/EPT-based efforts~\cite{azab2016skee, proskurin2020xmp, lvd20}. Furthermore, \sys allows developers to select specific objects for additional bound checking during data transfer. Combining SFI on the basis of PKS-based protection will realize finer isolation granularity at the cost of performance.

\PP{Policy Generation}
\label{sec-policy}
As a fundamental compartmentalization mechanism, \sys can support developer-defined policy flexibly. The LKM isolation use case deploys a relatively simple policy based on boundary analysis, shared data analysis, and security check analysis. More complex policies may further improve kernel security, but it is still an open problem to generate compartmentalization policy automatically. Basically, exploring the optimal policy can be reduced to the Partition Problem, which is NP-Complete~\cite{np03}. Some attempts have been made on bare-metal systems~\cite{aces18, tian23ec}. However, scaling the complex static analysis to a huge kernel like Linux is extremely hard. $\mu$SCOPE~\cite{roessler2021muscope} proposed another way based on dynamic analysis but cannot guarantee the soundness of the result. Combining static and dynamic analysis for policy generation is promising~\cite{ortega2022flexc}, and we expect type-based dependence analysis~\cite{lu2023practical} can provide a practical solution, which is one of our ongoing works.

\PP{Performance Optimization}
Though \sys shows acceptable overhead even for multiple kernel compartments. Its performance can be further optimized through some engineering efforts. One possible option is to eliminate redundant checks during compartment communication. For example, we can create a fast path for switches that occurs frequently in a short term once the two compartments pass the first check and establish trust for communication temporarily.

\section{Related Work}
\label{sec8}
\PP{Microkernels}
As opposed to the monolithic architecture, microkernels~\cite{microkernel95, klein2009sel4, gu2020harmonizing, redleaf20} maintain only the core functionality in the kernel space to minimize the attack surface. The reduced codebase facilitates formal verification~\cite{klein2009sel4}. More recently, some efforts even explored the development of microkernels in safe languages, such as RedLeaf~\cite{redleaf20} written in Rust. However, a cost coming with these security benefits is its low performance, especially the IPC overhead between isolated components~\cite{gu2020harmonizing}. Besides, the shift from traditional monolithic kernels to microkernels requires complete system redesign, while \sys enhances the monolithic kernel security with minor engineering effort through compartmentalization.

\PP{SFI-based Approaches}
Software fault isolation (SFI)~\cite{sfi93} inserts security checks during compilation time to regulate all accesses within specific domain boundaries. Following this basic idea, BGI~\cite{bgi09} manages an access control list to isolate Windows drivers. LXFI~\cite{mao2011software} further enforces kernel API integrity based on programmers' annotations. Nevertheless, the heavy checks impose a significant performance overhead. What's worse, as the number of domains increases and cross-domain interactions become more frequent, the performance of SFI-based approaches degrades sharply~\cite{sfiOverhead10}. For example, BGI induces over 30\% throughput loss for isolating many memory blocks. Benefiting from PKS-based permission validation, \sys avoids the significant overhead caused by software checks.

\PP{PT Switching and Virtualization-based Isolation}
Since the page table (PT) is a critical structure in charge of address translation and permission validation, PT switching is another well-explored mechanism for isolation. There are several works using different PTs to construct different memory views for userspace applications, such as lwC~\cite{lwc16} and SMV~\cite{smv16}. While representative studies for kernel isolation are Nooks~\cite{Nooks03}, SIDE~\cite{side13}, and SKEE~\cite{azab2016skee}, all suffer from significant overhead caused by updating PT and flushing TLB. With the development of virtualization technology, researchers further introduced the hypervisor to facilitate EPT switching and enhance security~\cite{xiong2011huko, virtuos13, secpod15, lxd19, lvd20, huang2022ksplit}. In contrast, BULKHEAD does not need the additional privilege layer as TCB, instead a lightweight in-kernel monitor enforces comprehensive security invariants.

\PP{Hardware-based Isolation}
Besides Intel PKS, researchers have explored various hardware features for isolation~\cite{dautenhahn2015nested, cho2017dynamic, MANES2018130DIKernel, sfitag23, mckee2022preventing, Capacity23}. Nested Kernel~\cite{dautenhahn2015nested} utilizes the WP (Write-Protect bit) mechanism of the x86-64 hardware for intra-kernel privilege separation. Hilps~\cite{cho2017dynamic} leverages the TxSZ mechanism of AArch64 to build a secure kernel domain by dynamic virtual address range adjustment. However, both approaches only support isolation between two domains. DIKernel~\cite{MANES2018130DIKernel} enforces isolation between the core kernel and extensions with the ARM Memory Domain Access Control mechanism, which has been deprecated recently. Memory Tagging Extension (MTE) and Pointer Authentication (PA) are recent features on ARM, which have been used for in-process compartmentalization~\cite{Capacity23} and kernel compartmentalization~\cite{mckee2022preventing} but show significant performance overhead. Works like Mondrix~\cite{witchel2005mondrix}, CODOMs~\cite{codoms14}, CHERI~\cite{watson2015cheri}, and SecureCells~\cite{bhattacharyya2023securecells} facilitate compartmentalization through newly designed hardware architecture. Compared to these efforts, \sys is based on the real available commodity hardware feature, which is more compatible and practical. Moreover, it breaks the hardware limitation of PKS to support unlimited compartments.

\PP{Other System Compartmentalization}
Compartmentalization has been explored as a principled defense against the myriad possible faults in software other than the monolithic kernel. Wedge~\cite{wedge08} splits complex applications into fine-grained, least-privilege compartments. SOAAP~\cite{soaap15} allows programmers to reason about application compartmentalization using source code annotations. Glamdring~\cite{Glamdring17} uses annotations and static analysis to partition applications for SGX. In addition to user-space applications, there is also a series of works on compartmentalizing embedded systems~\cite{aces18, opec22, tian23ec, tian23com2} and libOS~\cite{flexos22, sung2020intra, CubicleOS21}. For instance, ACES~\cite{aces18} enforces developer-specified policies with the MPU hardware feature for embedded system compartmentalization. EC~\cite{tian23ec} further provides a comprehensive and automatic compartmentalization toolchain for Real-Time Operating Systems (RTOSs) and bare-metal firmware. FlexOS~\cite{flexos22} specializes in the isolation strategy of a libOS at compilation/deployment time instead of design time, and enforces strategies with multiple hardware and software protection mechanisms. In contrast, \sys targets the monolithic kernel, which is more challenging due to its large codebase and privileged environment.

\section{Conclusion}
\label{sec9}

In this paper, we present \sys, a secure, scalable, and efficient kernel compartmentalization approach using a novel application of PKS. It guarantees bi-directional isolation between compartments and introduces a lightweight in-kernel monitor to enforce security-critical invariants, especially compartment interface integrity against confused deputy attacks. Besides, a locality-aware two-level scheme can provide unlimited compartments for scalability. We implement a prototype system for automated LKM isolation and extensive evaluations show that \sys incurs negligible performance overhead on real-world applications.

\section*{Acknowledgment}

We sincerely appreciate the anonymous reviewers for their insightful comments. Authors from Nanjing University were supported in part by NSFC under Grant Nos. 61772266, 61431008. Yinggang Guo was also partly supported by a scholarship from the Graduate School of Nanjing University and by the funding from the University of Minnesota. Weiheng Bai and Kangjie Lu were supported in part by NSF awards CNS-1815621, CNS1931208, CNS2045478, CNS-2106771, and CNS-2154989. Any opinions, findings, conclusions or recommendations expressed in this material are those of the authors and do not necessarily reflect the views of NSF.

\bibliographystyle{sty/IEEEtranS.bst}
\footnotesize
\setlength{\bibsep}{3pt}
\bibliography{p,conf}

\begin{thebibliography}{100}
\providecommand{\url}[1]{#1}
\csname url@samestyle\endcsname
\providecommand{\newblock}{\relax}
\providecommand{\bibinfo}[2]{#2}
\providecommand{\BIBentrySTDinterwordspacing}{\spaceskip=0pt\relax}
\providecommand{\BIBentryALTinterwordstretchfactor}{4}
\providecommand{\BIBentryALTinterwordspacing}{\spaceskip=\fontdimen2\font plus
\BIBentryALTinterwordstretchfactor\fontdimen3\font minus \fontdimen4\font\relax}
\providecommand{\BIBforeignlanguage}[2]{{%
\expandafter\ifx\csname l@#1\endcsname\relax
\typeout{** WARNING: IEEEtran.bst: No hyphenation pattern has been}%
\typeout{** loaded for the language `#1'. Using the pattern for}%
\typeout{** the default language instead.}%
\else
\language=\csname l@#1\endcsname
\fi
#2}}
\providecommand{\BIBdecl}{\relax}
\BIBdecl

\bibitem{CVE23Linux}
\BIBentryALTinterwordspacing
SecurityScorecard. (2024, Apr.) Linux kernel vulnerabilities. [Online]. Available: \url{https://www.cvedetails.com/product/47/Linux-Linux-Kernel.html?vendor_id=33}
\BIBentrySTDinterwordspacing

\bibitem{mckee2022preventing}
\BIBentryALTinterwordspacing
D.~McKee, Y.~Giannaris, C.~O. Perez, H.~Shrobe, M.~Payer, H.~Okhravi, and N.~Burow, ``Preventing kernel hacks with hakc,'' in \emph{Proceedings 2022 Network and Distributed System Security Symposium. NDSS}, vol.~22, 2022, pp. 1--17. [Online]. Available: \url{https://doi.org/10.14722/ndss.2022.24026}
\BIBentrySTDinterwordspacing

\bibitem{roessler2021muscope}
\BIBentryALTinterwordspacing
N.~Roessler, L.~Atayde, I.~Palmer, D.~McKee, J.~Pandey, V.~P. Kemerlis, M.~Payer, A.~Bates, J.~M. Smith, A.~DeHon, and N.~Dautenhahn, ``$\mu$scope: A methodology for analyzing least-privilege compartmentalization in large software artifacts,'' in \emph{24th International Symposium on Research in Attacks, Intrusions and Defenses}, ser. RAID '21.\hskip 1em plus 0.5em minus 0.4em\relax New York, NY, USA: Association for Computing Machinery, 2021, p. 296–311. [Online]. Available: \url{https://doi.org/10.1145/3471621.3471839}
\BIBentrySTDinterwordspacing

\bibitem{saltzer1975protection}
\BIBentryALTinterwordspacing
J.~H. Saltzer and M.~D. Schroeder, ``The protection of information in computer systems,'' \emph{Proceedings of the IEEE}, vol.~63, no.~9, pp. 1278--1308, 1975. [Online]. Available: \url{https://ieeexplore.ieee.org/abstract/document/1451869}
\BIBentrySTDinterwordspacing

\bibitem{Lefeuvre2023}
\BIBentryALTinterwordspacing
H.~Lefeuvre, V.-A. B{\u{a} }doiu, Y.~Chen, F.~Huici, N.~Dautenhahn, and P.~Olivier, ``Assessing the impact of interface vulnerabilities in compartmentalized software,'' in \emph{Proceedings 2023 Network and Distributed System Security Symposium}.\hskip 1em plus 0.5em minus 0.4em\relax Internet Society, 2023. [Online]. Available: \url{https://doi.org/10.14722%2Fndss.2023.24117}
\BIBentrySTDinterwordspacing

\bibitem{microkernel95}
\BIBentryALTinterwordspacing
J.~Liedtke, ``On micro-kernel construction,'' in \emph{Proceedings of the Fifteenth ACM Symposium on Operating Systems Principles}, ser. SOSP '95.\hskip 1em plus 0.5em minus 0.4em\relax New York, NY, USA: Association for Computing Machinery, 1995, p. 237–250. [Online]. Available: \url{https://doi.org/10.1145/224056.224075}
\BIBentrySTDinterwordspacing

\bibitem{klein2009sel4}
\BIBentryALTinterwordspacing
G.~Klein, K.~Elphinstone, G.~Heiser, J.~Andronick, D.~Cock, P.~Derrin, D.~Elkaduwe, K.~Engelhardt, R.~Kolanski, M.~Norrish, T.~Sewell, H.~Tuch, and S.~Winwood, ``sel4: formal verification of an os kernel,'' in \emph{Proceedings of the ACM SIGOPS 22nd Symposium on Operating Systems Principles}, ser. SOSP '09.\hskip 1em plus 0.5em minus 0.4em\relax New York, NY, USA: Association for Computing Machinery, 2009, p. 207–220. [Online]. Available: \url{https://doi.org/10.1145/1629575.1629596}
\BIBentrySTDinterwordspacing

\bibitem{gu2020harmonizing}
\BIBentryALTinterwordspacing
J.~Gu, X.~Wu, W.~Li, N.~Liu, Z.~Mi, Y.~Xia, and H.~Chen, ``Harmonizing performance and isolation in microkernels with efficient intra-kernel isolation and communication,'' in \emph{2020 USENIX Annual Technical Conference (USENIX ATC 20)}, 2020, pp. 401--417. [Online]. Available: \url{https://www.usenix.org/conference/atc20/presentation/gu}
\BIBentrySTDinterwordspacing

\bibitem{redleaf20}
\BIBentryALTinterwordspacing
V.~Narayanan, T.~Huang, D.~Detweiler, D.~Appel, Z.~Li, G.~Zellweger, and A.~Burtsev, ``{RedLeaf}: Isolation and communication in a safe operating system,'' in \emph{14th USENIX Symposium on Operating Systems Design and Implementation (OSDI 20)}.\hskip 1em plus 0.5em minus 0.4em\relax USENIX Association, Nov. 2020, pp. 21--39. [Online]. Available: \url{https://www.usenix.org/conference/osdi20/presentation/narayanan-vikram}
\BIBentrySTDinterwordspacing

\bibitem{sfi93}
\BIBentryALTinterwordspacing
R.~Wahbe, S.~Lucco, T.~E. Anderson, and S.~L. Graham, ``Efficient software-based fault isolation,'' \emph{SIGOPS Oper. Syst. Rev.}, vol.~27, no.~5, p. 203–216, Dec. 1993. [Online]. Available: \url{https://doi.org/10.1145/173668.168635}
\BIBentrySTDinterwordspacing

\bibitem{xfi06}
\BIBentryALTinterwordspacing
U.~Erlingsson, M.~Abadi, M.~Vrable, M.~Budiu, and G.~C. Necula, ``Xfi: Software guards for system address spaces,'' in \emph{Proceedings of the 7th Symposium on Operating Systems Design and Implementation}, ser. OSDI '06.\hskip 1em plus 0.5em minus 0.4em\relax USA: USENIX Association, 2006, p. 75–88. [Online]. Available: \url{https://www.usenix.org/conference/osdi-06/xfi-software-guards-system-address-spaces}
\BIBentrySTDinterwordspacing

\bibitem{bgi09}
\BIBentryALTinterwordspacing
M.~Castro, M.~Costa, J.-P. Martin, M.~Peinado, P.~Akritidis, A.~Donnelly, P.~Barham, and R.~Black, ``Fast byte-granularity software fault isolation,'' in \emph{Proceedings of the ACM SIGOPS 22nd Symposium on Operating Systems Principles}, ser. SOSP '09.\hskip 1em plus 0.5em minus 0.4em\relax New York, NY, USA: Association for Computing Machinery, 2009, p. 45–58. [Online]. Available: \url{https://doi.org/10.1145/1629575.1629581}
\BIBentrySTDinterwordspacing

\bibitem{mao2011software}
\BIBentryALTinterwordspacing
Y.~Mao, H.~Chen, D.~Zhou, X.~Wang, N.~Zeldovich, and M.~F. Kaashoek, ``Software fault isolation with api integrity and multi-principal modules,'' in \emph{Proceedings of the Twenty-Third ACM Symposium on Operating Systems Principles}, ser. SOSP '11.\hskip 1em plus 0.5em minus 0.4em\relax New York, NY, USA: Association for Computing Machinery, 2011, p. 115–128. [Online]. Available: \url{https://doi.org/10.1145/2043556.2043568}
\BIBentrySTDinterwordspacing

\bibitem{xiong2011huko}
\BIBentryALTinterwordspacing
X.~Xiong, D.~Tian, P.~Liu \emph{et~al.}, ``Practical protection of kernel integrity for commodity os from untrusted extensions.'' in \emph{NDSS}, vol.~11, 2011. [Online]. Available: \url{https://www.ndss-symposium.org/ndss2011/practical-protection-of-kernel-integrity-for-commodity-os-from-untrusted-extensions/}
\BIBentrySTDinterwordspacing

\bibitem{virtuos13}
\BIBentryALTinterwordspacing
R.~Nikolaev and G.~Back, ``Virtuos: An operating system with kernel virtualization,'' in \emph{Proceedings of the Twenty-Fourth ACM Symposium on Operating Systems Principles}, ser. SOSP '13.\hskip 1em plus 0.5em minus 0.4em\relax New York, NY, USA: Association for Computing Machinery, 2013, p. 116–132. [Online]. Available: \url{https://doi.org/10.1145/2517349.2522719}
\BIBentrySTDinterwordspacing

\bibitem{secpod15}
\BIBentryALTinterwordspacing
X.~Wang, Y.~Chen, Z.~Wang, Y.~Qi, and Y.~Zhou, ``{SecPod}: a framework for virtualization-based security systems,'' in \emph{2015 USENIX Annual Technical Conference (USENIX ATC 15)}.\hskip 1em plus 0.5em minus 0.4em\relax Santa Clara, CA: USENIX Association, Jul. 2015, pp. 347--360. [Online]. Available: \url{https://www.usenix.org/conference/atc15/technical-session/presentation/wang-xiaoguang}
\BIBentrySTDinterwordspacing

\bibitem{lxd19}
\BIBentryALTinterwordspacing
V.~Narayanan, A.~Balasubramanian, C.~Jacobsen, S.~Spall, S.~Bauer, M.~Quigley, A.~Hussain, A.~Younis, J.~Shen, M.~Bhattacharyya, and A.~Burtsev, ``{LXDs}: Towards isolation of kernel subsystems,'' in \emph{2019 USENIX Annual Technical Conference (USENIX ATC 19)}.\hskip 1em plus 0.5em minus 0.4em\relax Renton, WA: USENIX Association, Jul. 2019, pp. 269--284. [Online]. Available: \url{https://www.usenix.org/conference/atc19/presentation/narayanan}
\BIBentrySTDinterwordspacing

\bibitem{lvd20}
\BIBentryALTinterwordspacing
V.~Narayanan, Y.~Huang, G.~Tan, T.~Jaeger, and A.~Burtsev, ``Lightweight kernel isolation with virtualization and vm functions,'' in \emph{Proceedings of the 16th ACM SIGPLAN/SIGOPS International Conference on Virtual Execution Environments}, ser. VEE '20.\hskip 1em plus 0.5em minus 0.4em\relax New York, NY, USA: Association for Computing Machinery, 2020, p. 157–171. [Online]. Available: \url{https://doi.org/10.1145/3381052.3381328}
\BIBentrySTDinterwordspacing

\bibitem{huang2022ksplit}
\BIBentryALTinterwordspacing
Y.~Huang, V.~Narayanan, D.~Detweiler, K.~Huang, G.~Tan, T.~Jaeger, and A.~Burtsev, ``{KSplit}: Automating device driver isolation,'' in \emph{16th USENIX Symposium on Operating Systems Design and Implementation (OSDI 22)}.\hskip 1em plus 0.5em minus 0.4em\relax Carlsbad, CA: USENIX Association, Jul. 2022, pp. 613--631. [Online]. Available: \url{https://www.usenix.org/conference/osdi22/presentation/huang-yongzhe}
\BIBentrySTDinterwordspacing

\bibitem{dautenhahn2015nested}
\BIBentryALTinterwordspacing
N.~Dautenhahn, T.~Kasampalis, W.~Dietz, J.~Criswell, and V.~Adve, ``Nested kernel: An operating system architecture for intra-kernel privilege separation,'' in \emph{Proceedings of the Twentieth International Conference on Architectural Support for Programming Languages and Operating Systems}, 2015, pp. 191--206. [Online]. Available: \url{https://doi.org/10.1145/2694344.2694386}
\BIBentrySTDinterwordspacing

\bibitem{cho2017dynamic}
\BIBentryALTinterwordspacing
Y.~Cho, D.~Kwon, H.~Yi, and Y.~Paek, ``Dynamic virtual address range adjustment for intra-level privilege separation on arm,'' in \emph{NDSS}, 2017. [Online]. Available: \url{https://doi.org/10.14722/NDSS.2017.23024}
\BIBentrySTDinterwordspacing

\bibitem{MANES2018130DIKernel}
\BIBentryALTinterwordspacing
V.~J. Manès, D.~Jang, C.~Ryu, and B.~B. Kang, ``Domain isolated kernel: A lightweight sandbox for untrusted kernel extensions,'' \emph{Computers \& Security}, vol.~74, pp. 130--143, 2018. [Online]. Available: \url{https://www.sciencedirect.com/science/article/pii/S0167404818300282}
\BIBentrySTDinterwordspacing

\bibitem{iskios21}
\BIBentryALTinterwordspacing
S.~Gravani, M.~Hedayati, J.~Criswell, and M.~L. Scott, ``Fast intra-kernel isolation and security with iskios,'' in \emph{Proceedings of the 24th International Symposium on Research in Attacks, Intrusions and Defenses}, ser. RAID '21.\hskip 1em plus 0.5em minus 0.4em\relax New York, NY, USA: Association for Computing Machinery, 2021, p. 119–134. [Online]. Available: \url{https://doi.org/10.1145/3471621.3471849}
\BIBentrySTDinterwordspacing

\bibitem{syzkaller}
\BIBentryALTinterwordspacing
Google. (2024, Apr.) syzbot reported kernel bugs. [Online]. Available: \url{https://syzkaller.appspot.com/}
\BIBentrySTDinterwordspacing

\bibitem{minibox14bi}
\BIBentryALTinterwordspacing
Y.~Li, J.~McCune, J.~Newsome, A.~Perrig, B.~Baker, and W.~Drewry, ``{MiniBox}: A {Two-Way} sandbox for x86 native code,'' in \emph{2014 USENIX Annual Technical Conference (USENIX ATC 14)}.\hskip 1em plus 0.5em minus 0.4em\relax Philadelphia, PA: USENIX Association, Jun. 2014, pp. 409--420. [Online]. Available: \url{https://www.usenix.org/conference/atc14/technical-sessions/presentation/li_yanlin}
\BIBentrySTDinterwordspacing

\bibitem{sgxlock22}
\BIBentryALTinterwordspacing
Y.~Chen, J.~Li, G.~Xu, Y.~Zhou, Z.~Wang, C.~Wang, and K.~Ren, ``{SGXLock}: Towards efficiently establishing mutual distrust between host application and enclave for {SGX},'' in \emph{31st USENIX Security Symposium (USENIX Security 22)}.\hskip 1em plus 0.5em minus 0.4em\relax Boston, MA: USENIX Association, Aug. 2022, pp. 4129--4146. [Online]. Available: \url{https://www.usenix.org/conference/usenixsecurity22/presentation/chen-yuan}
\BIBentrySTDinterwordspacing

\bibitem{hotos23trust}
\BIBentryALTinterwordspacing
C.~Castes, A.~Ghosn, N.~S. Kalani, Y.~Qian, M.~Kogias, M.~Payer, and E.~Bugnion, ``Creating trust by abolishing hierarchies,'' in \emph{Proceedings of the 19th Workshop on Hot Topics in Operating Systems}, ser. HOTOS '23.\hskip 1em plus 0.5em minus 0.4em\relax New York, NY, USA: Association for Computing Machinery, 2023, p. 231–238. [Online]. Available: \url{https://doi.org/10.1145/3593856.3595900}
\BIBentrySTDinterwordspacing

\bibitem{civscope23}
\BIBentryALTinterwordspacing
Y.~Chien, V.-A. B\u{a}doiu, Y.~Yang, Y.~Huo, K.~Kaoudis, H.~Lefeuvre, P.~Olivier, and N.~Dautenhahn, ``Civscope: Analyzing potential memory corruption bugs in compartment interfaces,'' in \emph{Proceedings of the 1st Workshop on Kernel Isolation, Safety and Verification}, ser. KISV '23.\hskip 1em plus 0.5em minus 0.4em\relax New York, NY, USA: Association for Computing Machinery, 2023, p. 33–40. [Online]. Available: \url{https://doi.org/10.1145/3625275.3625399}
\BIBentrySTDinterwordspacing

\bibitem{hotos23interface}
\BIBentryALTinterwordspacing
A.~Burtsev, V.~Narayanan, Y.~Huang, K.~Huang, G.~Tan, and T.~Jaeger, ``Evolving operating system kernels towards secure kernel-driver interfaces,'' in \emph{Proceedings of the 19th Workshop on Hot Topics in Operating Systems}, ser. HOTOS '23.\hskip 1em plus 0.5em minus 0.4em\relax New York, NY, USA: Association for Computing Machinery, 2023, p. 166–173. [Online]. Available: \url{https://doi.org/10.1145/3593856.3595914}
\BIBentrySTDinterwordspacing

\bibitem{intel23manual}
\BIBentryALTinterwordspacing
Intel. (2023, Dec.) Intel 64 and ia-32 architectures software developer manuals. [Online]. Available: \url{https://www.intel.com/content/www/us/en/developer/articles/technical/intel-sdm.html}
\BIBentrySTDinterwordspacing

\bibitem{pitfall20}
\BIBentryALTinterwordspacing
R.~J. Connor, T.~McDaniel, J.~M. Smith, and M.~Schuchard, ``{PKU} pitfalls: Attacks on {PKU-based} memory isolation systems,'' in \emph{29th USENIX Security Symposium (USENIX Security 20)}.\hskip 1em plus 0.5em minus 0.4em\relax USENIX Association, Aug. 2020, pp. 1409--1426. [Online]. Available: \url{https://www.usenix.org/conference/usenixsecurity20/presentation/connor}
\BIBentrySTDinterwordspacing

\bibitem{voulimeneas2022you}
\BIBentryALTinterwordspacing
A.~Voulimeneas, J.~Vinck, R.~Mechelinck, and S.~Volckaert, ``You shall not (by)pass! practical, secure, and fast pku-based sandboxing,'' in \emph{Proceedings of the Seventeenth European Conference on Computer Systems}, ser. EuroSys '22.\hskip 1em plus 0.5em minus 0.4em\relax New York, NY, USA: Association for Computing Machinery, 2022, p. 266–282. [Online]. Available: \url{https://doi.org/10.1145/3492321.3519560}
\BIBentrySTDinterwordspacing

\bibitem{schrammel2022jenny}
\BIBentryALTinterwordspacing
D.~Schrammel, S.~Weiser, R.~Sadek, and S.~Mangard, ``Jenny: Securing syscalls for {PKU-based} memory isolation systems,'' in \emph{31st USENIX Security Symposium (USENIX Security 22)}.\hskip 1em plus 0.5em minus 0.4em\relax Boston, MA: USENIX Association, Aug. 2022, pp. 936--952. [Online]. Available: \url{https://www.usenix.org/conference/usenixsecurity22/presentation/schrammel}
\BIBentrySTDinterwordspacing

\bibitem{KASLR}
\BIBentryALTinterwordspacing
J.~Edge. (2013, Oct.) Kernel address space layout randomization. [Online]. Available: \url{https://lwn.net/Articles/569635/}
\BIBentrySTDinterwordspacing

\bibitem{phoronix}
P.~Media, ``{Phoronix test suites: Open-Source, Automated Benchmarking},'' \url{https://www.phoronix-test-suite.com/}, 2023.

\bibitem{ab}
T.~A.~S. Foundation, ``{Apache HTTP server benchmarking tool},'' \url{https://httpd.apache.org/docs/2.4/programs/ab.html}, 2023.

\bibitem{proskurin2020xmp}
\BIBentryALTinterwordspacing
S.~Proskurin, M.~Momeu, S.~Ghavamnia, V.~P. Kemerlis, and M.~Polychronakis, ``xmp: Selective memory protection for kernel and user space,'' in \emph{2020 IEEE Symposium on Security and Privacy (SP)}, 2020, pp. 563--577. [Online]. Available: \url{https://ieeexplore.ieee.org/abstract/document/9152671}
\BIBentrySTDinterwordspacing

\bibitem{azab2016skee}
\BIBentryALTinterwordspacing
A.~M. Azab, K.~Swidowski, R.~Bhutkar, J.~Ma, W.~Shen, R.~Wang, and P.~Ning, ``Skee: A lightweight secure kernel-level execution environment for arm,'' in \emph{NDSS}, vol.~16, 2016, pp. 21--24. [Online]. Available: \url{https://doi.org/10.14722/NDSS.2016.23009}
\BIBentrySTDinterwordspacing

\bibitem{watson2015cheri}
\BIBentryALTinterwordspacing
R.~N. Watson, J.~Woodruff, P.~G. Neumann, S.~W. Moore, J.~Anderson, D.~Chisnall, N.~Dave, B.~Davis, K.~Gudka, B.~Laurie, S.~J. Murdoch, R.~Norton, M.~Roe, S.~Son, and M.~Vadera, ``Cheri: A hybrid capability-system architecture for scalable software compartmentalization,'' in \emph{2015 IEEE Symposium on Security and Privacy}, 2015, pp. 20--37. [Online]. Available: \url{https://doi.org/10.1109/SP.2015.9}
\BIBentrySTDinterwordspacing

\bibitem{bhattacharyya2023securecells}
\BIBentryALTinterwordspacing
A.~Bhattacharyya, F.~Hofhammer, Y.~Li, S.~Gupta, A.~S{\'a}nchez~Mar{\'\i}n, B.~Falsafi, and M.~Payer, ``Securecells: A secure compartmentalized architecture,'' in \emph{44th IEEE Symposium on Security and Privacy}.\hskip 1em plus 0.5em minus 0.4em\relax Los Alamitos, CA, USA: IEEE Computer Society, may 2023, pp. 2921--2939. [Online]. Available: \url{https://doi.ieeecomputersociety.org/10.1109/SP46215.2023.00125}
\BIBentrySTDinterwordspacing

\bibitem{dope23}
\BIBentryALTinterwordspacing
L.~Maar, M.~Schwarzl, F.~Rauscher, D.~Gruss, and S.~Mangard, ``Dope: Domain protection enforcement with pks,'' in \emph{Proceedings of the 39th Annual Computer Security Applications Conference}, ser. ACSAC '23.\hskip 1em plus 0.5em minus 0.4em\relax New York, NY, USA: Association for Computing Machinery, 2023, p. 662–676. [Online]. Available: \url{https://doi.org/10.1145/3627106.3627113}
\BIBentrySTDinterwordspacing

\bibitem{hypsec19}
\BIBentryALTinterwordspacing
S.-W. Li, J.~S. Koh, and J.~Nieh, ``Protecting cloud virtual machines from hypervisor and host operating system exploits,'' in \emph{28th USENIX Security Symposium (USENIX Security 19)}.\hskip 1em plus 0.5em minus 0.4em\relax Santa Clara, CA: USENIX Association, Aug. 2019, pp. 1357--1374. [Online]. Available: \url{https://www.usenix.org/conference/usenixsecurity19/presentation/li-shih-wei}
\BIBentrySTDinterwordspacing

\bibitem{cloud20}
\BIBentryALTinterwordspacing
Z.~Mi, D.~Li, H.~Chen, B.~Zang, and H.~Guan, ``(mostly) exitless {VM} protection from untrusted hypervisor through disaggregated nested virtualization,'' in \emph{29th USENIX Security Symposium (USENIX Security 20)}.\hskip 1em plus 0.5em minus 0.4em\relax USENIX Association, Aug. 2020, pp. 1695--1712. [Online]. Available: \url{https://www.usenix.org/conference/usenixsecurity20/presentation/mi}
\BIBentrySTDinterwordspacing

\bibitem{biIso23}
\BIBentryALTinterwordspacing
J.~Park, S.~Kang, S.~Lee, T.~Kim, J.~Park, Y.~Kwon, and J.~Huh, ``Hardware hardened sandbox enclaves for trusted serverless computing,'' \emph{ACM Trans. Archit. Code Optim.}, nov 2023, just Accepted. [Online]. Available: \url{https://doi.org/10.1145/3632954}
\BIBentrySTDinterwordspacing

\bibitem{lin2022dirtycred}
\BIBentryALTinterwordspacing
Z.~Lin, Y.~Wu, and X.~Xing, ``Dirtycred: Escalating privilege in linux kernel,'' in \emph{Proceedings of the 2022 ACM SIGSAC Conference on Computer and Communications Security}, ser. CCS '22.\hskip 1em plus 0.5em minus 0.4em\relax New York, NY, USA: Association for Computing Machinery, 2022, p. 1963–1976. [Online]. Available: \url{https://doi.org/10.1145/3548606.3560585}
\BIBentrySTDinterwordspacing

\bibitem{kcofi14}
\BIBentryALTinterwordspacing
J.~Criswell, N.~Dautenhahn, and V.~Adve, ``Kcofi: Complete control-flow integrity for commodity operating system kernels,'' in \emph{2014 IEEE Symposium on Security and Privacy}, 2014, pp. 292--307. [Online]. Available: \url{https://doi.org/10.1109/SP.2014.26}
\BIBentrySTDinterwordspacing

\bibitem{dop16}
\BIBentryALTinterwordspacing
H.~Hu, S.~Shinde, S.~Adrian, Z.~L. Chua, P.~Saxena, and Z.~Liang, ``Data-oriented programming: On the expressiveness of non-control data attacks,'' in \emph{2016 IEEE Symposium on Security and Privacy (SP)}, 2016, pp. 969--986. [Online]. Available: \url{https://ieeexplore.ieee.org/abstract/document/7546545}
\BIBentrySTDinterwordspacing

\bibitem{disclosure16}
\BIBentryALTinterwordspacing
J.~Gionta, W.~Enck, and P.~Larsen, ``Preventing kernel code-reuse attacks through disclosure resistant code diversification,'' in \emph{2016 IEEE Conference on Communications and Network Security (CNS)}, 2016, pp. 189--197. [Online]. Available: \url{https://ieeexplore.ieee.org/abstract/document/7860485}
\BIBentrySTDinterwordspacing

\bibitem{vmPerformance15}
\BIBentryALTinterwordspacing
W.~Felter, A.~Ferreira, R.~Rajamony, and J.~Rubio, ``An updated performance comparison of virtual machines and linux containers,'' in \emph{2015 IEEE International Symposium on Performance Analysis of Systems and Software (ISPASS)}, 2015, pp. 171--172. [Online]. Available: \url{https://ieeexplore.ieee.org/servlet/opac?punumber=7093633}
\BIBentrySTDinterwordspacing

\bibitem{performance21}
\BIBentryALTinterwordspacing
V.~van Rijn and J.~S. Rellermeyer, ``A fresh look at the architecture and performance of contemporary isolation platforms,'' in \emph{Proceedings of the 22nd International Middleware Conference}, ser. Middleware '21.\hskip 1em plus 0.5em minus 0.4em\relax New York, NY, USA: Association for Computing Machinery, 2021, p. 323–335. [Online]. Available: \url{https://doi.org/10.1145/3464298.3493404}
\BIBentrySTDinterwordspacing

\bibitem{vahldiek2019erim}
\BIBentryALTinterwordspacing
A.~Vahldiek-Oberwagner, E.~Elnikety, N.~O. Duarte, M.~Sammler, P.~Druschel, and D.~Garg, ``{ERIM}: Secure, efficient in-process isolation with protection keys ({{{{{MPK}}}}}),'' in \emph{28th USENIX Security Symposium (USENIX Security 19)}.\hskip 1em plus 0.5em minus 0.4em\relax Santa Clara, CA: USENIX Association, Aug. 2019, pp. 1221--1238. [Online]. Available: \url{https://www.usenix.org/conference/usenixsecurity19/presentation/vahldiek-oberwagner}
\BIBentrySTDinterwordspacing

\bibitem{hedayati2019hodor}
\BIBentryALTinterwordspacing
M.~Hedayati, S.~Gravani, E.~Johnson, J.~Criswell, M.~L. Scott, K.~Shen, and M.~Marty, ``Hodor:$\{$Intra-Process$\}$ isolation for $\{$High-Throughput$\}$ data plane libraries,'' in \emph{2019 USENIX Annual Technical Conference (USENIX ATC 19)}, 2019, pp. 489--504. [Online]. Available: \url{https://www.usenix.org/conference/atc19/presentation/hedayati}
\BIBentrySTDinterwordspacing

\bibitem{kuzuno2022kdpm}
\BIBentryALTinterwordspacing
H.~Kuzuno and T.~Yamauchi, ``Kdpm: Kernel data protection mechanism using a memory protection key,'' in \emph{Advances in Information and Computer Security}.\hskip 1em plus 0.5em minus 0.4em\relax Cham: Springer International Publishing, 2022, pp. 66--84. [Online]. Available: \url{https://doi.org/10.1007/978-3-031-15255-9_4}
\BIBentrySTDinterwordspacing

\bibitem{moat24}
\BIBentryALTinterwordspacing
H.~Lu, S.~Wang, Y.~Wu, W.~He, and F.~Zhang, ``{MOAT}: Towards safe {BPF} kernel extension,'' in \emph{33rd USENIX Security Symposium (USENIX Security 24)}.\hskip 1em plus 0.5em minus 0.4em\relax Philadelphia, PA: USENIX Association, Aug. 2024, pp. 1153--1170. [Online]. Available: \url{https://www.usenix.org/conference/usenixsecurity24/presentation/lu-hongyi}
\BIBentrySTDinterwordspacing

\bibitem{pksjos}
\BIBentryALTinterwordspacing
H.~LI, J.-Y. GU, Y.-B. XIA, B.-Y. ZANG, and H.-B. CHEN, ``Memory isolation mechanism of ebpf based on pks hardware feature,'' \emph{Journal of Software}, vol.~34, no.~12, p. 5921, 2023. [Online]. Available: \url{https://www.jos.org.cn/josen/article/abstract/6762}
\BIBentrySTDinterwordspacing

\bibitem{suh2003aegis}
\BIBentryALTinterwordspacing
G.~E. Suh, D.~Clarke, B.~Gassend, M.~Van~Dijk, and S.~Devadas, ``Aegis: Architecture for tamper-evident and tamper-resistant processing,'' in \emph{ACM International Conference on Supercomputing 25th Anniversary Volume}, 2003, pp. 357--368. [Online]. Available: \url{https://doi.org/10.1145/782814.782838}
\BIBentrySTDinterwordspacing

\bibitem{wilkins2013uefi}
\BIBentryALTinterwordspacing
R.~Wilkins and B.~Richardson, ``Uefi secure boot in modern computer security solutions,'' in \emph{UEFI forum}, 2013, pp. 1--10. [Online]. Available: \url{https://api.semanticscholar.org/CorpusID:14326971}
\BIBentrySTDinterwordspacing

\bibitem{yitbarek2017cold}
\BIBentryALTinterwordspacing
S.~F. Yitbarek, M.~T. Aga, R.~Das, and T.~Austin, ``Cold boot attacks are still hot: Security analysis of memory scramblers in modern processors,'' in \emph{2017 IEEE International Symposium on High Performance Computer Architecture (HPCA)}.\hskip 1em plus 0.5em minus 0.4em\relax IEEE, 2017, pp. 313--324. [Online]. Available: \url{https://doi.org/10.1109/HPCA.2017.10}
\BIBentrySTDinterwordspacing

\bibitem{mutlu2019rowhammer}
\BIBentryALTinterwordspacing
O.~Mutlu and J.~S. Kim, ``Rowhammer: A retrospective,'' \emph{IEEE Transactions on Computer-Aided Design of Integrated Circuits and Systems}, vol.~39, no.~8, pp. 1555--1571, 2019. [Online]. Available: \url{https://doi.org/10.1109/TCAD.2019.2915318}
\BIBentrySTDinterwordspacing

\bibitem{spectre19}
\BIBentryALTinterwordspacing
P.~Kocher, J.~Horn, A.~Fogh, D.~Genkin, D.~Gruss, W.~Haas, M.~Hamburg, M.~Lipp, S.~Mangard, T.~Prescher, M.~Schwarz, and Y.~Yarom, ``Spectre attacks: Exploiting speculative execution,'' in \emph{2019 IEEE Symposium on Security and Privacy (SP)}, 2019, pp. 1--19. [Online]. Available: \url{https://doi.org/10.1109/SP.2019.00002}
\BIBentrySTDinterwordspacing

\bibitem{meltdown18}
\BIBentryALTinterwordspacing
M.~Lipp, M.~Schwarz, D.~Gruss, T.~Prescher, W.~Haas, A.~Fogh, J.~Horn, S.~Mangard, P.~Kocher, D.~Genkin, Y.~Yarom, and M.~Hamburg, ``Meltdown: Reading kernel memory from user space,'' in \emph{Proceedings of the 27th USENIX Conference on Security Symposium}, ser. SEC'18.\hskip 1em plus 0.5em minus 0.4em\relax USA: USENIX Association, 2018, p. 973–990. [Online]. Available: \url{https://doi.org/10.1145/3357033}
\BIBentrySTDinterwordspacing

\bibitem{fan2023isa}
\BIBentryALTinterwordspacing
S.~Fan, Z.~Hua, Y.~Xia, H.~Chen, and B.~Zang, ``Isa-grid: Architecture of fine-grained privilege control for instructions and registers,'' in \emph{Proceedings of the 50th Annual International Symposium on Computer Architecture}, ser. ISCA '23.\hskip 1em plus 0.5em minus 0.4em\relax New York, NY, USA: Association for Computing Machinery, 2023. [Online]. Available: \url{https://doi.org/10.1145/3579371.3589050}
\BIBentrySTDinterwordspacing

\bibitem{wu2022dancing}
\BIBentryALTinterwordspacing
C.~Wu, M.~Xie, Z.~Wang, Y.~Zhang, K.~Lu, X.~Zhang, Y.~Lai, Y.~Kang, M.~Yang, and T.~Li, ``Dancing with wolves: An intra-process isolation technique with privileged hardware,'' \emph{IEEE Transactions on Dependable and Secure Computing}, vol.~20, no.~3, pp. 1959--1978, 2023. [Online]. Available: \url{https://ieeexplore.ieee.org/abstract/document/9760152}
\BIBentrySTDinterwordspacing

\bibitem{kernelXOM19}
\BIBentryALTinterwordspacing
M.~Pomonis, T.~Petsios, A.~D. Keromytis, M.~Polychronakis, and V.~P. Kemerlis, ``Kernel protection against just-in-time code reuse,'' \emph{ACM Trans. Priv. Secur.}, vol.~22, no.~1, jan 2019. [Online]. Available: \url{https://doi.org/10.1145/3277592}
\BIBentrySTDinterwordspacing

\bibitem{rop12}
\BIBentryALTinterwordspacing
R.~Roemer, E.~Buchanan, H.~Shacham, and S.~Savage, ``Return-oriented programming: Systems, languages, and applications,'' \emph{ACM Trans. Inf. Syst. Secur.}, vol.~15, no.~1, mar 2012. [Online]. Available: \url{https://doi.org/10.1145/2133375.2133377}
\BIBentrySTDinterwordspacing

\bibitem{finecfi18}
\BIBentryALTinterwordspacing
J.~Li, X.~Tong, F.~Zhang, and J.~Ma, ``Fine-cfi: Fine-grained control-flow integrity for operating system kernels,'' \emph{IEEE Transactions on Information Forensics and Security}, vol.~13, no.~6, pp. 1535--1550, 2018. [Online]. Available: \url{https://ieeexplore.ieee.org/abstract/document/8269390}
\BIBentrySTDinterwordspacing

\bibitem{mlta19lu}
\BIBentryALTinterwordspacing
K.~Lu and H.~Hu, ``Where does it go? refining indirect-call targets with multi-layer type analysis,'' in \emph{Proceedings of the 2019 ACM SIGSAC Conference on Computer and Communications Security}, ser. CCS '19.\hskip 1em plus 0.5em minus 0.4em\relax New York, NY, USA: Association for Computing Machinery, 2019, p. 1867–1881. [Online]. Available: \url{https://doi.org/10.1145/3319535.3354244}
\BIBentrySTDinterwordspacing

\bibitem{typro22}
\BIBentryALTinterwordspacing
M.~Bauer, I.~Grishchenko, and C.~Rossow, ``Typro: Forward cfi for c-style indirect function calls using type propagation,'' in \emph{Proceedings of the 38th Annual Computer Security Applications Conference}, ser. ACSAC '22.\hskip 1em plus 0.5em minus 0.4em\relax New York, NY, USA: Association for Computing Machinery, 2022, p. 346–360. [Online]. Available: \url{https://doi.org/10.1145/3564625.3564627}
\BIBentrySTDinterwordspacing

\bibitem{schrammel2020donky}
\BIBentryALTinterwordspacing
D.~Schrammel, S.~Weiser, S.~Steinegger, M.~Schwarzl, M.~Schwarz, S.~Mangard, and D.~Gruss, ``Donky: Domain keys {\textendash} efficient {In-Process} isolation for {RISC-V} and x86,'' in \emph{29th USENIX Security Symposium (USENIX Security 20)}.\hskip 1em plus 0.5em minus 0.4em\relax USENIX Association, Aug. 2020, pp. 1677--1694. [Online]. Available: \url{https://www.usenix.org/conference/usenixsecurity20/presentation/schrammel}
\BIBentrySTDinterwordspacing

\bibitem{hardwarepk20}
\BIBentryALTinterwordspacing
Y.~Xu, C.~Ye, Y.~Solihin, and X.~Shen, ``Hardware-based domain virtualization for intra-process isolation of persistent memory objects,'' in \emph{2020 ACM/IEEE 47th Annual International Symposium on Computer Architecture (ISCA)}, 2020, pp. 680--692. [Online]. Available: \url{https://doi.org/10.1109/ISCA45697.2020.00062}
\BIBentrySTDinterwordspacing

\bibitem{libmpk19}
\BIBentryALTinterwordspacing
S.~Park, S.~Lee, W.~Xu, H.~Moon, and T.~Kim, ``libmpk: Software abstraction for intel memory protection keys (intel {MPK}),'' in \emph{2019 USENIX Annual Technical Conference (USENIX ATC 19)}.\hskip 1em plus 0.5em minus 0.4em\relax Renton, WA: USENIX Association, Jul. 2019, pp. 241--254. [Online]. Available: \url{https://www.usenix.org/conference/atc19/presentation/park-soyeon}
\BIBentrySTDinterwordspacing

\bibitem{gu2022epk}
\BIBentryALTinterwordspacing
J.~Gu, H.~Li, W.~Li, Y.~Xia, and H.~Chen, ``{EPK}: Scalable and efficient memory protection keys,'' in \emph{2022 USENIX Annual Technical Conference (USENIX ATC 22)}.\hskip 1em plus 0.5em minus 0.4em\relax Carlsbad, CA: USENIX Association, Jul. 2022, pp. 609--624. [Online]. Available: \url{https://www.usenix.org/conference/atc22/presentation/gu-jinyu}
\BIBentrySTDinterwordspacing

\bibitem{yuan2023vdom}
\BIBentryALTinterwordspacing
Z.~Yuan, S.~Hong, R.~Chang, Y.~Zhou, W.~Shen, and K.~Ren, ``Vdom: Fast and unlimited virtual domains on multiple architectures,'' in \emph{Proceedings of the 28th ACM International Conference on Architectural Support for Programming Languages and Operating Systems, Volume 2}, ser. ASPLOS 2023.\hskip 1em plus 0.5em minus 0.4em\relax New York, NY, USA: Association for Computing Machinery, 2023, p. 905–919. [Online]. Available: \url{https://doi.org/10.1145/3575693.3575735}
\BIBentrySTDinterwordspacing

\bibitem{formal22}
\BIBentryALTinterwordspacing
Y.~Guo, Z.~Wang, B.~Zhong, and Q.~Zeng, ``Formal modeling and security analysis for intra-level privilege separation,'' in \emph{Proceedings of the 38th Annual Computer Security Applications Conference}, ser. ACSAC '22.\hskip 1em plus 0.5em minus 0.4em\relax New York, NY, USA: Association for Computing Machinery, 2022, p. 88–101. [Online]. Available: \url{https://doi.org/10.1145/3564625.3567984}
\BIBentrySTDinterwordspacing

\bibitem{lu19missing-check}
\BIBentryALTinterwordspacing
K.~Lu, A.~Pakki, and Q.~Wu, ``Detecting {Missing-Check} bugs via semantic- and {Context-Aware} criticalness and constraints inferences,'' in \emph{28th USENIX Security Symposium (USENIX Security 19)}.\hskip 1em plus 0.5em minus 0.4em\relax Santa Clara, CA: USENIX Association, Aug. 2019, pp. 1769--1786. [Online]. Available: \url{https://www.usenix.org/conference/usenixsecurity19/presentation/lu}
\BIBentrySTDinterwordspacing

\bibitem{svf16}
\BIBentryALTinterwordspacing
Y.~Sui and J.~Xue, ``Svf: Interprocedural static value-flow analysis in llvm,'' in \emph{Proceedings of the 25th International Conference on Compiler Construction}, ser. CC 2016.\hskip 1em plus 0.5em minus 0.4em\relax New York, NY, USA: Association for Computing Machinery, 2016, p. 265–266. [Online]. Available: \url{https://doi.org/10.1145/2892208.2892235}
\BIBentrySTDinterwordspacing

\bibitem{dma21}
\BIBentryALTinterwordspacing
J.-J. Bai, T.~Li, K.~Lu, and S.-M. Hu, ``Static detection of unsafe {DMA} accesses in device drivers,'' in \emph{30th USENIX Security Symposium (USENIX Security 21)}.\hskip 1em plus 0.5em minus 0.4em\relax USENIX Association, Aug. 2021, pp. 1629--1645. [Online]. Available: \url{https://www.usenix.org/conference/usenixsecurity21/presentation/bai}
\BIBentrySTDinterwordspacing

\bibitem{lu2023practical}
\BIBentryALTinterwordspacing
K.~Lu, ``Practical program modularization with type-based dependence analysis,'' in \emph{2023 IEEE Symposium on Security and Privacy (SP)}.\hskip 1em plus 0.5em minus 0.4em\relax Los Alamitos, CA, USA: IEEE Computer Society, may 2023, pp. 1610--1624. [Online]. Available: \url{https://doi.ieeecomputersociety.org/10.1109/SP46215.2023.00092}
\BIBentrySTDinterwordspacing

\bibitem{mcvoy1996lmbench}
\BIBentryALTinterwordspacing
L.~W. McVoy, C.~Staelin \emph{et~al.}, ``lmbench: Portable tools for performance analysis.'' in \emph{USENIX annual technical conference}.\hskip 1em plus 0.5em minus 0.4em\relax San Diego, CA, USA, 1996, pp. 279--294. [Online]. Available: \url{https://www.usenix.org/conference/usenix-1996-annual-technical-conference/lmbench-portable-tools-performance-analysis}
\BIBentrySTDinterwordspacing

\bibitem{sok19benchmark}
\BIBentryALTinterwordspacing
E.~van~der Kouwe, G.~Heiser, D.~Andriesse, H.~Bos, and C.~Giuffrida, ``Sok: Benchmarking flaws in systems security,'' in \emph{2019 IEEE European Symposium on Security and Privacy (EuroS\&P)}, 2019, pp. 310--325. [Online]. Available: \url{https://ieeexplore.ieee.org/abstract/document/8806739}
\BIBentrySTDinterwordspacing

\bibitem{np03}
\BIBentryALTinterwordspacing
E.~D. Demaine, S.~Hohenberger, and D.~Liben-Nowell, ``Tetris is hard, even to approximate,'' in \emph{Computing and Combinatorics}, T.~Warnow and B.~Zhu, Eds.\hskip 1em plus 0.5em minus 0.4em\relax Berlin, Heidelberg: Springer Berlin Heidelberg, 2003, pp. 351--363. [Online]. Available: \url{https://erikdemaine.org/papers/Tetris_TR2002/}
\BIBentrySTDinterwordspacing

\bibitem{aces18}
\BIBentryALTinterwordspacing
A.~A. Clements, N.~S. Almakhdhub, S.~Bagchi, and M.~Payer, ``{ACES}: Automatic compartments for embedded systems,'' in \emph{27th USENIX Security Symposium (USENIX Security 18)}.\hskip 1em plus 0.5em minus 0.4em\relax Baltimore, MD: USENIX Association, Aug. 2018, pp. 65--82. [Online]. Available: \url{https://www.usenix.org/conference/usenixsecurity18/presentation/clements}
\BIBentrySTDinterwordspacing

\bibitem{tian23ec}
\BIBentryALTinterwordspacing
A.~Khan, D.~Xu, and D.~J. Tian, ``Ec: Embedded systems compartmentalization via intra-kernel isolation,'' in \emph{2023 IEEE Symposium on Security and Privacy (SP)}, 2023, pp. 2990--3007. [Online]. Available: \url{https://ieeexplore.ieee.org/document/10179285}
\BIBentrySTDinterwordspacing

\bibitem{ortega2022flexc}
\BIBentryALTinterwordspacing
C.~P. Ortega, ``Flexc: Flexible compartmentalization through automatic policy generation,'' Master's thesis, Massachusetts Institute of Technology, 2022. [Online]. Available: \url{https://dspace.mit.edu/bitstream/handle/1721.1/144506/Ortega-cortegap-meng-eecs-2022-thesis.pdf?sequence=1}
\BIBentrySTDinterwordspacing

\bibitem{sfiOverhead10}
\BIBentryALTinterwordspacing
D.~Sehr, R.~Muth, C.~Biffle, V.~Khimenko, E.~Pasko, K.~Schimpf, B.~Yee, and B.~Chen, ``Adapting software fault isolation to contemporary cpu architectures,'' in \emph{Proceedings of the 19th USENIX Conference on Security}, ser. USENIX Security'10.\hskip 1em plus 0.5em minus 0.4em\relax USA: USENIX Association, 2010, p.~1. [Online]. Available: \url{https://www.usenix.org/conference/usenixsecurity10/adapting-software-fault-isolation-contemporary-cpu-architectures}
\BIBentrySTDinterwordspacing

\bibitem{lwc16}
\BIBentryALTinterwordspacing
J.~Litton, A.~Vahldiek-Oberwagner, E.~Elnikety, D.~Garg, B.~Bhattacharjee, and P.~Druschel, ``{Light-Weight} contexts: An {OS} abstraction for safety and performance,'' in \emph{12th USENIX Symposium on Operating Systems Design and Implementation (OSDI 16)}.\hskip 1em plus 0.5em minus 0.4em\relax Savannah, GA: USENIX Association, Nov. 2016, pp. 49--64. [Online]. Available: \url{https://www.usenix.org/conference/osdi16/technical-sessions/presentation/litton}
\BIBentrySTDinterwordspacing

\bibitem{smv16}
\BIBentryALTinterwordspacing
T.~C.-H. Hsu, K.~Hoffman, P.~Eugster, and M.~Payer, ``Enforcing least privilege memory views for multithreaded applications,'' in \emph{Proceedings of the 2016 ACM SIGSAC Conference on Computer and Communications Security}, ser. CCS '16.\hskip 1em plus 0.5em minus 0.4em\relax New York, NY, USA: Association for Computing Machinery, 2016, p. 393–405. [Online]. Available: \url{https://doi.org/10.1145/2976749.2978327}
\BIBentrySTDinterwordspacing

\bibitem{Nooks03}
\BIBentryALTinterwordspacing
M.~M. Swift, B.~N. Bershad, and H.~M. Levy, ``Improving the reliability of commodity operating systems,'' \emph{SIGOPS Oper. Syst. Rev.}, vol.~37, no.~5, p. 207–222, oct 2003. [Online]. Available: \url{https://doi.org/10.1145/1165389.945466}
\BIBentrySTDinterwordspacing

\bibitem{side13}
\BIBentryALTinterwordspacing
Y.~Sun and T.-c. Chiueh, ``Side: Isolated and efficient execution of unmodified device drivers,'' in \emph{2013 43rd Annual IEEE/IFIP International Conference on Dependable Systems and Networks (DSN)}, 2013, pp. 1--12. [Online]. Available: \url{https://doi.org/10.1109/DSN.2013.6575348}
\BIBentrySTDinterwordspacing

\bibitem{sfitag23}
\BIBentryALTinterwordspacing
J.~Seo, J.~You, Y.~Cho, Y.~Cho, D.~Kwon, and Y.~Paek, ``Sfitag: Efficient software fault isolation with memory tagging for arm kernel extensions,'' in \emph{Proceedings of the 2023 ACM Asia Conference on Computer and Communications Security}, ser. ASIA CCS '23.\hskip 1em plus 0.5em minus 0.4em\relax New York, NY, USA: Association for Computing Machinery, 2023, p. 469–480. [Online]. Available: \url{https://doi.org/10.1145/3579856.3590341}
\BIBentrySTDinterwordspacing

\bibitem{Capacity23}
\BIBentryALTinterwordspacing
K.~Dinh~Duy, K.~Cho, T.~Noh, and H.~Lee, ``Capacity: Cryptographically-enforced in-process capabilities for modern arm architectures,'' in \emph{Proceedings of the 2023 ACM SIGSAC Conference on Computer and Communications Security}, ser. CCS '23.\hskip 1em plus 0.5em minus 0.4em\relax New York, NY, USA: Association for Computing Machinery, 2023, p. 874–888. [Online]. Available: \url{https://doi.org/10.1145/3576915.3623079}
\BIBentrySTDinterwordspacing

\bibitem{witchel2005mondrix}
\BIBentryALTinterwordspacing
E.~Witchel, J.~Rhee, and K.~Asanovi{\'c}, ``Mondrix: Memory isolation for linux using mondriaan memory protection,'' in \emph{Proceedings of the twentieth ACM symposium on Operating systems principles}, 2005, pp. 31--44. [Online]. Available: \url{https://doi.org/10.1145/1095810.1095814}
\BIBentrySTDinterwordspacing

\bibitem{codoms14}
\BIBentryALTinterwordspacing
L.~Vilanova, M.~Ben-Yehuda, N.~Navarro, Y.~Etsion, and M.~Valero, ``Codoms: Protecting software with code-centric memory domains,'' in \emph{Proceeding of the 41st Annual International Symposium on Computer Architecuture}, ser. ISCA '14.\hskip 1em plus 0.5em minus 0.4em\relax IEEE Press, 2014, p. 469–480. [Online]. Available: \url{https://dl.acm.org/doi/abs/10.1145/2678373.2665741}
\BIBentrySTDinterwordspacing

\bibitem{wedge08}
\BIBentryALTinterwordspacing
A.~Bittau, P.~Marchenko, M.~Handley, and B.~Karp, ``Wedge: Splitting applications into reduced-privilege compartments,'' in \emph{Proceedings of the 5th USENIX Symposium on Networked Systems Design and Implementation}, ser. NSDI'08.\hskip 1em plus 0.5em minus 0.4em\relax USA: USENIX Association, 2008, p. 309–322. [Online]. Available: \url{https://www.usenix.org/conference/nsdi-08/wedge-splitting-applications-reduced-privilege-compartments}
\BIBentrySTDinterwordspacing

\bibitem{soaap15}
\BIBentryALTinterwordspacing
K.~Gudka, R.~N. Watson, J.~Anderson, D.~Chisnall, B.~Davis, B.~Laurie, I.~Marinos, P.~G. Neumann, and A.~Richardson, ``Clean application compartmentalization with soaap,'' in \emph{Proceedings of the 22nd ACM SIGSAC Conference on Computer and Communications Security}, ser. CCS '15.\hskip 1em plus 0.5em minus 0.4em\relax New York, NY, USA: Association for Computing Machinery, 2015, p. 1016–1031. [Online]. Available: \url{https://doi.org/10.1145/2810103.2813611}
\BIBentrySTDinterwordspacing

\bibitem{Glamdring17}
\BIBentryALTinterwordspacing
J.~Lind, C.~Priebe, D.~Muthukumaran, D.~O{\textquoteright}Keeffe, P.-L. Aublin, F.~Kelbert, T.~Reiher, D.~Goltzsche, D.~Eyers, R.~Kapitza, C.~Fetzer, and P.~Pietzuch, ``Glamdring: Automatic application partitioning for intel {SGX},'' in \emph{2017 USENIX Annual Technical Conference (USENIX ATC 17)}.\hskip 1em plus 0.5em minus 0.4em\relax Santa Clara, CA: USENIX Association, Jul. 2017, pp. 285--298. [Online]. Available: \url{https://www.usenix.org/conference/atc17/technical-sessions/presentation/lind}
\BIBentrySTDinterwordspacing

\bibitem{opec22}
\BIBentryALTinterwordspacing
X.~Zhou, J.~Li, W.~Zhang, Y.~Zhou, W.~Shen, and K.~Ren, ``Opec: operation-based security isolation for bare-metal embedded systems,'' in \emph{Proceedings of the Seventeenth European Conference on Computer Systems}, ser. EuroSys '22.\hskip 1em plus 0.5em minus 0.4em\relax New York, NY, USA: Association for Computing Machinery, 2022, p. 317–333. [Online]. Available: \url{https://doi.org/10.1145/3492321.3519573}
\BIBentrySTDinterwordspacing

\bibitem{tian23com2}
\BIBentryALTinterwordspacing
A.~Khan, D.~Xu, and D.~J. Tian, ``Low-cost privilege separation with compile time compartmentalization for embedded systems,'' in \emph{2023 IEEE Symposium on Security and Privacy (SP)}, 2023, pp. 3008--3025. [Online]. Available: \url{https://doi.org/10.1109/SP46215.2023.10179388}
\BIBentrySTDinterwordspacing

\bibitem{flexos22}
\BIBentryALTinterwordspacing
H.~Lefeuvre, V.-A. B\u{a}doiu, A.~Jung, S.~L. Teodorescu, S.~Rauch, F.~Huici, C.~Raiciu, and P.~Olivier, ``Flexos: Towards flexible os isolation,'' in \emph{Proceedings of the 27th ACM International Conference on Architectural Support for Programming Languages and Operating Systems}, ser. ASPLOS '22.\hskip 1em plus 0.5em minus 0.4em\relax New York, NY, USA: Association for Computing Machinery, 2022, p. 467–482. [Online]. Available: \url{https://doi.org/10.1145/3503222.3507759}
\BIBentrySTDinterwordspacing

\bibitem{sung2020intra}
\BIBentryALTinterwordspacing
M.~Sung, P.~Olivier, S.~Lankes, and B.~Ravindran, ``Intra-unikernel isolation with intel memory protection keys,'' in \emph{Proceedings of the 16th ACM SIGPLAN/SIGOPS International Conference on Virtual Execution Environments}, 2020, pp. 143--156. [Online]. Available: \url{https://doi.org/10.1145/3381052.3381326}
\BIBentrySTDinterwordspacing

\bibitem{CubicleOS21}
\BIBentryALTinterwordspacing
V.~A. Sartakov, L.~Vilanova, and P.~Pietzuch, ``Cubicleos: a library os with software componentisation for practical isolation,'' in \emph{Proceedings of the 26th ACM International Conference on Architectural Support for Programming Languages and Operating Systems}, ser. ASPLOS '21.\hskip 1em plus 0.5em minus 0.4em\relax New York, NY, USA: Association for Computing Machinery, 2021, p. 546–558. [Online]. Available: \url{https://doi.org/10.1145/3445814.3446731}
\BIBentrySTDinterwordspacing

\end{thebibliography}

\end{document}